\documentclass[journal,article,submit,moreauthors,pdftex]{{elsarticle}} 
\pdfoutput=1
\usepackage{bm}
\usepackage{indentfirst}
\usepackage{stfloats}
\usepackage{CJK}
\usepackage{amsmath}
\usepackage{graphicx}
\usepackage{subfigure}
\usepackage{diagbox}
\usepackage{booktabs}
\usepackage{textcomp,booktabs}
\usepackage{multirow}
\usepackage{colortbl}
\definecolor{sacolor1}{RGB}{140, 67, 46}
\definecolor{sacolor2}{RGB}{0, 0, 255}
\definecolor{sacolor3}{RGB}{255, 100, 0}
\definecolor{sacolor4}{RGB}{0, 255, 123}
\definecolor{sacolor5}{RGB}{164, 75, 155}
\definecolor{sacolor6}{RGB}{101, 174, 255}
\definecolor{sacolor7}{RGB}{118, 254, 172}
\definecolor{sacolor8}{RGB}{60, 91, 112}
\definecolor{sacolor9}{RGB}{255, 255, 0}
\definecolor{sacolor10}{RGB}{255, 255, 125}
\definecolor{sacolor11}{RGB}{255, 0, 255}
\definecolor{sacolor12}{RGB}{100, 0, 255}
\definecolor{sacolor13}{RGB}{0, 172, 254}
\definecolor{sacolor14}{RGB}{0, 255, 0}
\definecolor{sacolor15}{RGB}{171, 175, 80}
\definecolor{sacolor16}{RGB}{101, 193, 60}

\definecolor{unicolor1}{RGB}{192, 192, 192}
\definecolor{unicolor2}{RGB}{0, 255, 0}
\definecolor{unicolor3}{RGB}{0, 255, 255}
\definecolor{unicolor4}{RGB}{0, 128, 0}
\definecolor{unicolor5}{RGB}{255, 0, 255}
\definecolor{unicolor6}{RGB}{165, 82, 41}
\definecolor{unicolor7}{RGB}{128, 0, 128}
\definecolor{unicolor8}{RGB}{255, 0, 0}
\definecolor{unicolor9}{RGB}{255, 255, 0 }

\definecolor{jascolor1}{RGB}{255, 0, 0}
\definecolor{jascolor2}{RGB}{0, 0, 255}
\definecolor{jascolor3}{RGB}{0, 255, 0}
\definecolor{jascolor4}{RGB}{0, 0, 0}

\usepackage{array}
\usepackage{slashbox}

\begin{document}
\begin{frontmatter}	

\title{Hyperspectral Image Classification Based on Adaptive Sparse Deep Network}


\author[firstaddress]{Jingwen Yan}
\ead{jwyan@stu.edu.cn}

\author[firstaddress]{Zixin Xie}
\ead{18zxxie1@stu.edu.cn}

\author[firstaddress]{Jingyao Chen}
\ead{18jychen3@stu.edu.cn}

\author[firstaddress]{Yinan Liu}
\ead{13ynliu@stu.edu.cn}

\author[secondaddress]{Lei Liu \corref{cor1}}
\ead{wliulei@stu.edu.cn}

\cortext[cor1]{Corresponding author}
\address[firstaddress]{ Department of Electronic Engineering, Shantou University, Shantou 515063, China}
\address[secondaddress]{ Medical College, Shantou University, Shantou 515063, China;}

\begin{abstract}
Sparse model is widely used in hyperspectral image classification.
However, different of sparsity and regularization parameters has great influence on the classification results.
In this paper, a novel adaptive sparse deep network based on deep architecture is proposed, which can construct the optimal sparse representation and regularization parameters by deep network.
Firstly, a data flow graph is designed to represent each update iteration based on Alternating Direction Method of Multipliers (ADMM) algorithm.
Forward network and Back-Propagation network are deduced.
All parameters are updated by gradient descent in Back-Propagation.
Then we proposed an Adaptive Sparse Deep Network.
Comparing with several traditional classifiers or other algorithm for sparse model, experiment results indicate that our method achieves great improvement in HSI classification.
\end{abstract}


\end{frontmatter}

\section{Introduction}

Hyperspectral images (HSIs) have abundant spectral information and high spatial resolution, which has been widely used to scene classification, object detection and environmental monitoring, etc~\cite{Benediktsson2005Classification,Iyer2017Hyperspectral,ISI:000460321300028}.
In recent year, hyperspectral image classification has become a research hotspot in recent years and plays an important role in HSI analysis~\cite{Lin2018Recent}.
Multiple classification techniques have been attempted and achieved good performance~\cite{ISI:000477884900001}.

In recent years, sparse representation has attracted more and more attention from scholars.
It has the advantage of high accuracy and fast processing speed, and it does not need to make statistics and assumptions on sample distribution~\cite{Donoho2006Compressed}.
Therefore, it has been a powerful tool for signal processing and analysis, widely applied to many fields, such as data compression~\cite{Wang2017Sparse}, image restoration~\cite{Julien2007Sparse,Michael2006Image,ISI:000318016600030}, and face recognition ~\cite{ISI:000395476200008,John2009Robust,ISI:000399396400035}.
Recently, many classification model based on sparse representation have been proposed and have demonstrated superiority in hyperspectral image classification~\cite{ISI:000456936500022}.
In 2011, Chen et al. proposed a region-based sparse representation for hyperspectral image classification, rooted in the assumption that every hyperspectral pixel belonging to one class can be represented by a common sub-dictionary consisting of training samples from the same class~\cite{Yi2011Hyperspectral}.
In the experiment, in order to achieve the higher classification accuracy, the authors got the relative optimal values by setting up different parameter values (eg.sparsity level, weighting factor and neighborhood size).
In 2014, Zhang et al. proposed a method called non local weighed joint sparse representation classification (NLW-JSRC) to compensate for the deficiency of utilizing different weights for different neighboring pixels around the central test pixel in  joint sparsity model (JSM)~\cite{Zhang2017A,Baron2012Distributed}.
The NLW-JSRC method aims to ensure that the pixels which are similar to the central test pixel contribute more to the classification process, with a larger weight, and vice versa, which supports an improved hyperspectral image classification performance.
However, the authors manually set parameter values (eg.sparsity level and neighborhood size) for classification accuracy in the experiment.
In 2017, Fang et al proposed a multiple-feature-based adaptive sparse representation (MFASR) method for the hyperspectral images classification~\cite{Fang2017Hyperspectral}.
The MFASR utilizes an adaptive sparse representation to effectively exploit the correlations among four extracted features from the original HSI.
Finally the authors still need to set different parameter values for achieving the higher classification accuracy.
Therefore, for these hyperspectral image classification based on sparse representation, sparsity level is be paid more attention to~\cite{ChenhdaHyperspectral,Wei2016Hyperspectral}.

A hyperspectral pixel can be represented by a dictionary consisting of different classes extracted training samples from sparse samples in sparse representation classification model.
An unknown pixel can be expressed as a sparse vector whose a few nonzero entries correspond to the weight of the selected training samples.
The sparse vector can be recovered by solving a sparsity-constrained optimization problem ($\ell_{0}$ minimization problem), and the class label of the test hyperspectral pixels can be directly determined by the property of the recovered sparse vectors.
Therefore, how to solve the sparsity optimization problem is the key of sparse model~\cite{Zhang2017}.
Thera are many algorithms have been proposed and applied to solve the sparsity optimization problem including the greedy algorithms~\cite{ISI:000480665700027} and some $\ell-norm$ optimization algorithms~\cite{ISI:000457458000013}.
However, the classification results based on optimization algorithms depend on two parameters: the regularization parameter $\lambda$ and the sparsity level parameter $K$.
For the greedy algorithms, such as Orthogonal Matching Pursuit (OMP)~\cite{Davenport2010Analysis} and Subspace Pursuit (SP)~\cite{Dai2009Subspace}, they need to set up the appropriate $K$ in advance.
Even the now improved OMP algorithms were proposed.
Deanna Needell proposed the Regularized Orthogonal Matching Pursuit (ROMP) that combines the speed and ease of implementation of the greedy methods with the strong guarantees of the convex programming methods~\cite{Needell2010Signal}.
Meanwhile, for the pursuing efficiency in reconstructing sparse signals, Jian W et al. proposed the generalized OMP (gOMP) that is finished with much smaller number of iterations when compared to OMP~\cite{Jian2011Generalized}.
But parameter $K$ still need to be considered in advance.
For $\ell_{1}$ optimization algorithms, such as a Fast Iterative Shrinkage-Thresholding Algorithm (FISTA)~\cite{Beck2009A}, Bregman~\cite{Ye2011Split} and the framework of Sparse Reconstruction by Separable Approximation (SpaRSA)~\cite{Wright2009Sparse}, all these algorithms need to set regularization parameter $\lambda$ in advance when solving optimization problems.

The parameter values need to be set artificially and they are selected depending on producing better results by repeating the algorithm many times.
When the result is not met the target requirement, the parameters values will be adjusted again.
We can't test all the values and then to choose the best one, so that the theoretical optimal parameter values cannot be obtained and only the relative optimal parameter values can be selected.
The results of parameters values are set artificially and the non-adaptability or non-automation during the solution process, which limit the application of the sparse representation method.

To address the issues of the parameters values in the hyperspectral image classification based on sparse representation, Alternating Direction Method of Multipliers (ADMM) has been proposed and it is an efficient variable splitting algorithm with convergence guarantee.
It considers the augmented Lagrangian function of a given sparsity model, and splits variables into subgroups, which can be alternatively optimized by solving a few simply subproblems~\cite{Boyd2010Distributed}.
Although ADMM is generally efficient, it is not easy to determine the optimal parameters (e.g. penalty parameters) influencing the classification accuracy and speed in the sparse model.
Many applications have shown that penalty parameters which be chosen too large or too small can lead to a significant influence time of the solution or result of classification~\cite{1983,Fukushima1992Application,Spyridon1998A}.
Recently, Yang et al. proposed a Deep ADMM-Net for compressive sensing MRI, which greatly resolved parameter setting in this problem~\cite{YangADMM}.
The author firstly devised a network structure based on ADMM algorithm.
Related parameters are updated end-to-end using L-BFGS algorithm automatically rather than set up in advance.

\begin{figure}
	\centering
	\includegraphics[width=0.6\linewidth]{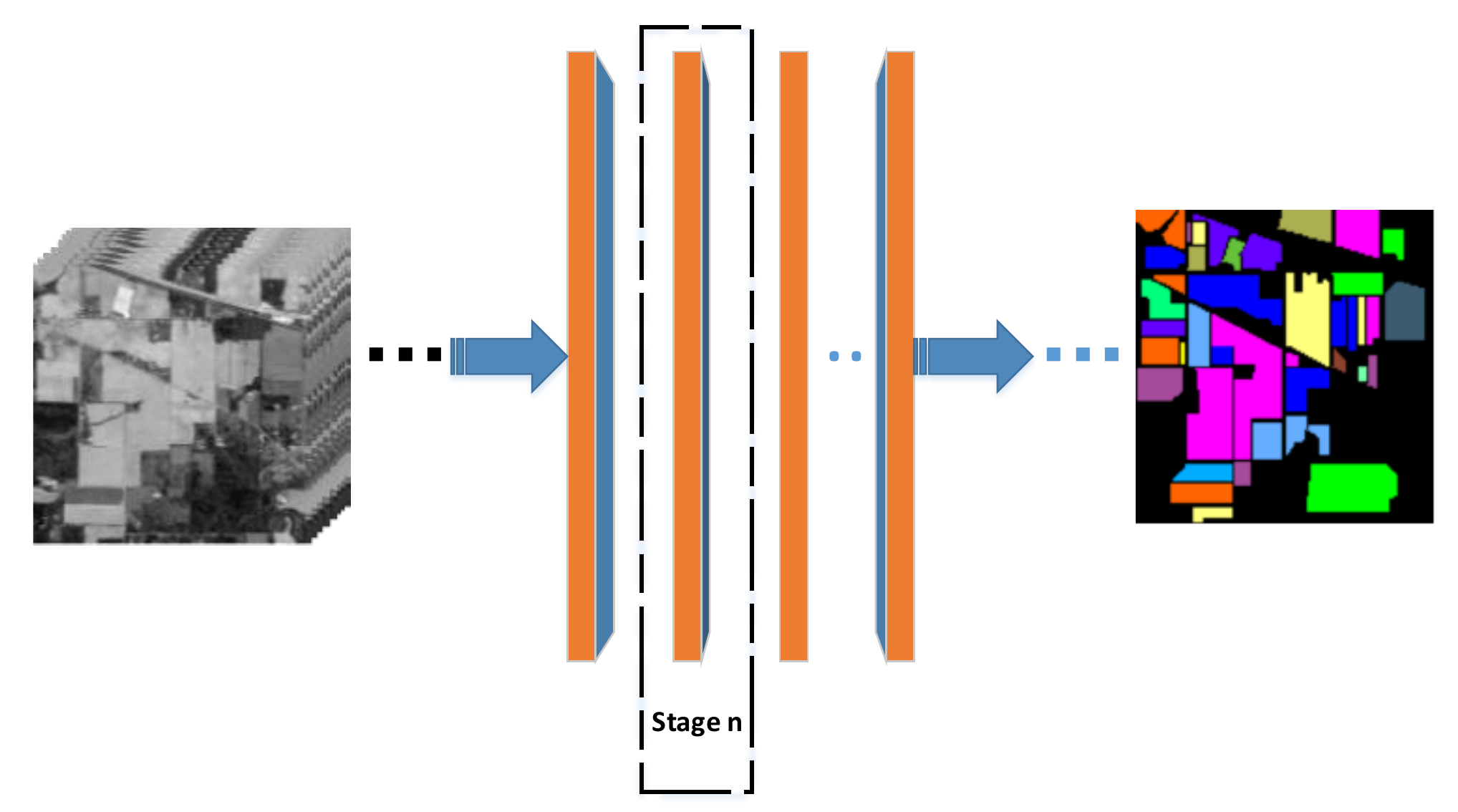}
	\caption{The whole data flow graph for the Adaptive Sparse Deep Network of hyperspectral sparse model.}
	\label{the whole data flow graph}
\end{figure}

In this paper, in order to construct the optimal parameters in a sparse model, we propose the Adaptive Sparse Deep Network based on a novel deep architecture.
Our work includes two parts.
Firstly, we introduce the iterative procedures in ADMM algorithm for optimizing a sparse model.
The data flow graph is shown in Fig.\ref{the whole data flow graph}.
It consists of multiple stages are each of which corresponds to an iteration in ADMM algorithm.
The proposed deep architecture is visible and not a black-box optimizers, different with a fully connected network or convolution network in deep  learning~\cite{Luo2018HSI,Slavkovikj2015Hyperspectral,ISI:000456936500044}.
Then, similar to the Feedforward Neural Network (FNN), the Adaptive Sparse Deep Network is divided into three parts, namely input layer, hidden layer and output layer.
Meanwhile, in the hidden layer, it has three types of operations, corresponding to different learnable parameters.
Finally, all parameters can be updated by gradient descent in Backward-Propagation.
An optimal sparse vector can be obtained by our method.
The optimal results of hyperspectral image classification can be got directly.

The remainder of this paper is structured as follows.
Section 2 reviews the sparse representation classification and hyperspectral sparse model on ADMM algorithm.
Section 3 proposes the Adaptive Sparse Deep Network for the hyperspectral image classification.
The experimental results of the proposed deep architecture with two hyperspectral images are given in Section 4.
Finally, Section 5 summarizes this paper and makes some closing remarks.

\begin{figure}[b]
	\centering
	\includegraphics[width=0.6\linewidth]{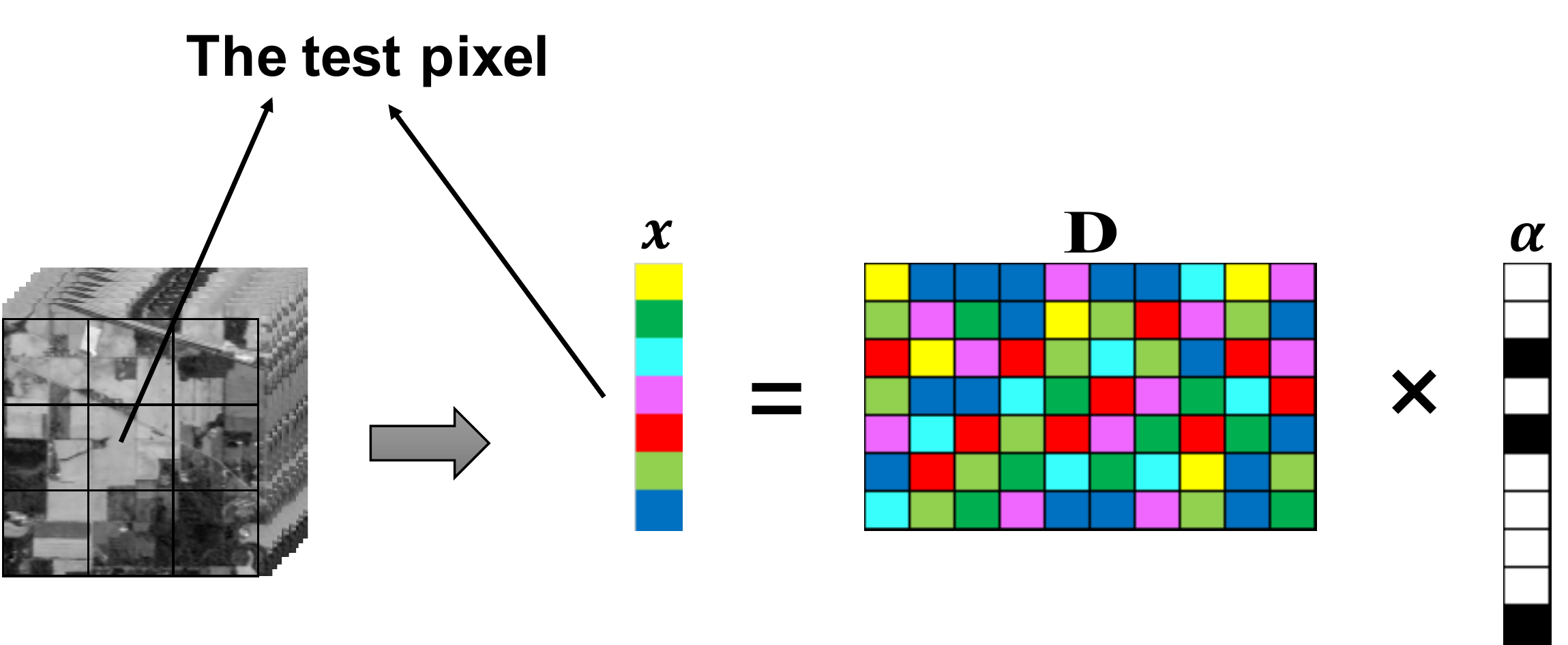}
	\caption{The Sparse Representation}
	\label{SR}
\end{figure}

\section{Hyperspectral sparse model based on the alternating direction methods of multipliers }

For a hyperspectral pixel $x \in R^{L}$, where L is the number of spectral bands, $x$ means the special line of a certain spatial location in HSI.
%
Given a dictionary $\bm{D= [{D_1},...,{D_i},...,{D_c}]} \in {R^{N \times M}}$, where $C$ is the number of the classes, $\bm{D_i} \in R^{L \times M_i}$ is a sub-dictionary, which is composed of $M_i$ pixels selected from the $ith$ class HSI.
According to sparse representation, a hyperspectral pixel can be represented by a linear combination of the dictionary $\bm{D}$.
As can been seen in Fig.\ref{SR}, the $x$ can be represented as follows:
\begin{equation}\label{eq:sr}
x = \bm{D}\alpha  + \varepsilon
\end{equation}

\noindent where $\alpha  \in {R^{M \times 1}}$ is sparse vector for the $x$ and $\varepsilon$ is an error residual item.
Here $\alpha$ can be obtained by solving following constrained optimization problem and hyperspectral sparse model is shown as follows:
\begin{equation}\label{eq:l0}
 \mathop {\arg \min }\limits_\alpha  {\left\| {x - \bm{D}\alpha } \right\|_2}{\rm{  }}\quad subject{\rm{ }}\ to \ {\rm{ }}{\left\| \alpha  \right\|_0} \le {K}
\end{equation}

\noindent where $K$ is a given upper bound on the sparsity level, which is equal to the upper bound number of nonzeros rows in $\alpha$.
Besides, for solving the problem (\ref{eq:l0}), the greedy algorithms can be used to solve the optimization problem.
However, the sparsity level parameter $K$ are needed to set up in advance.
Higher the sparsity $K$ will lead to higher computational cost and worse in the classification performance because misleading the dictionary atoms form the wrong classes to be selected.

The optimization problem (\ref{eq:l0}) also can be converted to a linear programming problem by replacing $\ell_{0}$ with $\ell_{1}$.
It is shown as follows:
\begin{equation}\label{eq:BP}
\arg \min {\left\| \alpha  \right\|_1}{\rm{  \quad  }} \ subject{\rm{ }} \ to \ {\rm{ }}\bm{D}\alpha  = x
\end{equation}

\noindent By adding the regularization parameter $\lambda$, the desired $\alpha$ is as sparse as possible and the error is as small as possible.
The problem (\ref{eq:BP}) can also be described as:
\begin{equation}\label{eq:l1}
\mathop {\arg \min }\limits_\alpha  \frac{1}{2}\left\| {x - \bm{D}\alpha } \right\|_2^2 + \lambda {\left\| \alpha  \right\|_1}
\end{equation}

There are many optimization algorithms for solving problem (\ref{eq:l1}), such as FISTA~\cite{Beck2009A}, Bregman iterative algorithm~\cite{Ye2011Split}, and SpaRSA~\cite{Wright2009Sparse}.
%
%
For all of the methods, the regularization parameter $\lambda$ provides a tradeoff between fidelity to the measurements and the sensitivity, which still need to set up in advance by human experience. 
%

In order to solve the issue of finding the optimal solutions in problem (\ref{eq:l1}), ADMM, a simple but powerful algorithm that is well suited to distributed convex optimization and takes the form of a decomposition-coordination procedure, in which the solutions to small local subproblems are coordinated to find a solution to a large global problem.
ADMM can be viewed as an attempt to blend the benefits of dual decomposition and augmented Lagrangian methods for constrained optimization~\cite{Boyd2010Distributed}.
Therefore, in this paper, this method is suitable for solving problem (\ref{eq:l1}).

By introducing auxiliary variables $z \in {R^{M \times 1}}$ , problem (\ref{eq:l1}) is equivalent to:
\begin{equation}\label{eq:origin}
\mathop {\arg \min }\limits_\alpha  \frac{1}{2}\left\| {x - \bm{D}\alpha } \right\|_2^2 + \lambda {\left\| z \right\|_1} \ \ {\rm{  subject \ to }}\ \alpha  - z = 0
\end{equation}

\noindent Its augmented Lagrangian function is:
\begin{equation}\label{eq:augmented Lagrangian}
L(\alpha ,z,u) = \frac{1}{2}\left\| {x - \bm{D}\alpha } \right\|_2^2 + \lambda {\left\| z \right\|_1} - y(z - \alpha ) +\frac{\rho }{2}\left\| {z - \alpha } \right\|_2^2
\end{equation}

\noindent where $y$ is Lagrangian multiplier and $\rho$ is penalty parameter.

For the sake of simplicity, formula (\ref{eq:augmented Lagrangian}) can also be written in the following form:
\begin{equation}\label{eq:simple}
L(\alpha ,z,u) = \frac{1}{2}\left\| {x - \bm{D}\alpha } \right\|_2^2 + \lambda {\left\| z \right\|_1} + \frac{\rho }{2}\left\| {z - \alpha  - u} \right\|_2^2 - \frac{\rho }{2}\left\| u \right\|_2^2
\end{equation}

\noindent here, let $u = \frac{y}{\rho}$.

In the sparse model, it commonly need to run the ADMM algorithm in dozens of iterations to get a optimal sparse vector.
Once $\alpha$ is obtained, the class of the hyperspectral pixel $x$ can be directly determined.
Comparing with the reconstruction results of each class, we classify $x$ by assigning it to the object class that minimizes the residual ${r_i}$:
\begin{equation}\label{eq:class}
\begin{array}{l}
class(x) = \mathop {\arg \min }\limits_{i = 1,...,C} {r_i}(x)\\
{\rm{             }} \qquad \quad \: \: \, = \mathop {\arg \min }\limits_{i = 1,...,C} {\left\| {x - {\bm{D_i}}{\alpha _i}} \right\|_2}
\end{array}
\end{equation}

Experience on applications has shown that the number of iterations depends significantly on the penalty parameter $\rho$.
If the fixed penalty parameter is chosen too small or too large, the solution time can increase significantly.
Just like sparsity level $K$ or regularization parameter $\lambda$ above, they are unknown and need to be predetermined.
After determining theirs value, the model is solved and the classification result is compared with the real labels.
If the result is not met, the parameters will be adjusted again, which results in the non-adaptability or non-automation of the solution process.

In the following section, some self-adaptive rules are applied for adjusting the penalty parameter and regularization parameter.
We can use the gradient descent by Back-Propagation to solve this problem.
Updating the corresponding parameters by the gradient computation, the self-adaptive parameters can be obtained.
In the next section, the detail of a deep architecture based on ADMM algorithm will be introduced.

\section{The Adaptive Sparse Deep Network}

Parameters setting are mostly depended on human experience, which is unsure that the obtained solution is the global optimal result.
Our work includes two parts.
Firstly, the Adaptive Sparse Deep Network (ASDN) is designed based on ADMM algorithm.
A deep data flow graph is shown in Fig.\ref{data flow graph}.
Each update iteration based on ADMM is corresponding to a stage in Fig.\ref{data flow graph}.
Secondly, the detail about updating the parameters by the gradient decent in the Back-Propagation will be introduced.
Clear network structure and data conduction relationships are shown in our methods.
In ASDN, all parameters are self-adaptively updated rather than manual setting.

\subsection{A Data Flow Graph of Updating Order of Parameters}

Data transfer and feedback are the basis of deep networks.
Firstly, a data flow graph is built, while the order of updating parameters are the key to it.
Here, problem (\ref{eq:simple}) can be solved by the following three subproblems based on ADMM algorithm:
\begin{equation}\label{eq:subproblems}
\left\{ \begin{array}{l}
{\alpha^{(n + 1)}}: = \mathop {\mathop {\arg \min }\limits_\alpha  (\frac{1}{2}\left\| {x - \bm{D}\alpha } \right\|_2^2 + \frac{\rho }{2}\left\| {{z^{(n)}} - \alpha  - {u^{(n)}}} \right\|_2^2)} \\
{z^{(n + 1)}}: = \mathop {\mathop {\arg \min }\limits_z (\lambda {{\left\| z \right\|}_1} + \frac{\rho }{2}\left\| {z - {\alpha ^{(n + 1)}} - {u^{(n)}}} \right\|_2^2)}  \\
{u^{(n + 1)}}: = \mathop {\arg \min }\limits_u (\frac{\rho }{2}\left\| {{z^{(n + 1)}} - {\alpha ^{(n + 1)}} - u} \right\|_2^2 - \frac{\rho }{2}\left\| u \right\|_2^2)   
\end{array} \right.
\end{equation}

\noindent Let $\eta = \frac{\lambda}{\rho}$, then the three subproblems have the following solutions:
\begin{equation}\label{eq:solutions}
\left\{ \begin{array}{l}
{\alpha ^{(n + 1)}} = (alpha)*{(\bm{{D^T}D} + \rho I)^{ - 1}}(\bm{{D^T}}x + \rho {z^{(n)}} - \rho {u^{(n)}})
{\rm{              \,+ }}(1 - alpha)*z{\rm{             }}\\
{z^{(n + 1)}} = S({\alpha ^{(n + 1)}} + {u^{(n)}},\eta )\\
{u^{(n + 1)}} = {u^{(n)}} + \tau ({\alpha ^{(n + 1)}} - {z^{(n + 1)}})
\end{array} \right.
\end{equation}

\noindent where $alpha$ is a relaxation parameter that can improve convergence in ADMM, $S( \cdot )$ is a nonlinear shrinkage function and $\tau$ denotes an update rate.
Therefore, in ADMM algorithm, the parameters updates as following order: the sparse vector ($\alpha$), then the non-linear vector ($z$) and finally the multiplier ($u$).

\begin{figure}
	\centering
	\includegraphics[width=1.0\linewidth]{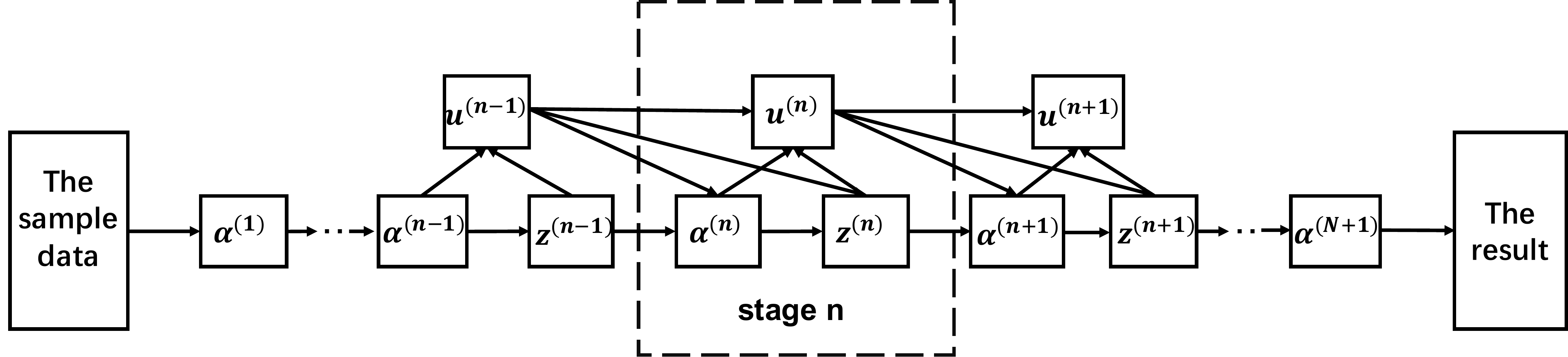}
	\caption{The data flow graph for the Adaptive Sparse Deep Network. The graph consists of three types of nodes: sparse vector ($\alpha$), non-linear vector ($z$) and multiplier ($u$).}
	\label{data flow graph}
\end{figure}

Based on the order of updating parameters, a deep data flow graph is devised which mapping the iteration of the ADMM algorithm.
As shown in Fig.\ref{data flow graph}, this graph consists of nodes corresponding to different operation in ADMM, and directed edges corresponding to the data flows between operations.
Each update (\ref{eq:subproblems}) is considered as a stage.
Similarity, the $nth$ iteration of ADMM algorithm corresponds to the $nth$ stage of the data flow graph.
In the $nth$ stage of the graph, there are three types of nodes mapped from three types of operation in ADMM, i.e., sparse vector operation (${\alpha ^{(n)}}$), nonlinear transform operation (${z^{(n)}}$) defined by $S( \cdot )$, and multiplier update operation (${{u^{(n)}}}$).
Therefore, the whole data flow graph corresponds well to each update iteration in ADMM.
Given an under-sampled data from HSIs, it flows over the graph and finally obtains the optimal sparse vector $\alpha$.
In the following subsections, the Adaptive Sparse Deep Network will be proposed, which is a deep architecture based on the data flow graph.

\subsection{The Architecture of the Adaptive Sparse Deep Network}

In this subsection, we propose a deep architecture dubbed Adaptive Sparse Deep Network.
It mainly consists three parts: input layer, hidden layer and output layer.
Meanwhile, in the hidden layer, it has three types of operations corresponding to different learnable parameters.
%

Sparsity operation: This operation obtains a sparse vector according to Eqn.(\ref{eq:subproblems}) and Eqn.(\ref{eq:solutions}).
Given ${z^{(n - 1)}}$ and ${{u^{(n - 1)}}}$, the output of the node is defined as:
\begin{equation}\label{eq:sparsity operation}
\begin{array}{l}
{\alpha ^{(n)}} = (alpha)*{(\bm{{D^T}D} + {\rho ^{(n)}}I)^{ - 1}}(\bm{{D^T}}x + {\rho ^{(n)}}{z^{(n - 1)}} - 
 {\rm{           }}{\rho ^{(n)}}{u^{(n - 1)}}) \\
\qquad \qquad + (1 - alpha)*{z^{(n - 1)}}
\end{array}
\end{equation}

\noindent where ${\rho ^{(n)}}$ are learnable parameters.
In the first stage $(n = 1)$,  ${z^{(0)}}$ and ${u^{(0)}}$ are initialized to zeros.
Therefore ${\alpha ^{(1)}} = (alpha) * {(\bm{{D^T}D} + {\rho ^{(1)}}I)^{ - 1}}(\bm{{D^T}}x)$.

Nonlinear transform operation: This operation performs nonlinear transform inspired by the ADMM algorithm in a sparse model in Eqn.(\ref{eq:subproblems}) and Eqn.(\ref{eq:solutions}).
Given ${\alpha ^{(n)}}$ and ${u^{(n - 1)}}$, the output of the node is defined as:
\begin{equation}\label{eq:nonlinear transform operation}
{z^{(n)}}: = S({\alpha ^{(n)}} + {u^{(n - 1)}},{\eta ^{(n)}})
\end{equation}

\noindent where ${\eta ^{(n)}}$ are learnable parameters.

Multiplier update operation: This operation is defined by the ADMM algorithm in Eqn.(\ref{eq:subproblems}) and Eqn.(\ref{eq:solutions}).
The output of this node in stage $n$ is defined as:
\begin{equation}\label{eq:multiplier update operation}
{u^{(n)}}: = {u^{(n - 1)}} + {\tau ^{(n)}}({\alpha ^{(n)}} - {z^{(n)}})
\end{equation}

\noindent where ${\tau ^{(n)}}$ are learnable parameters.

\begin{figure*}
	\centering
	\includegraphics[width=0.9\linewidth]{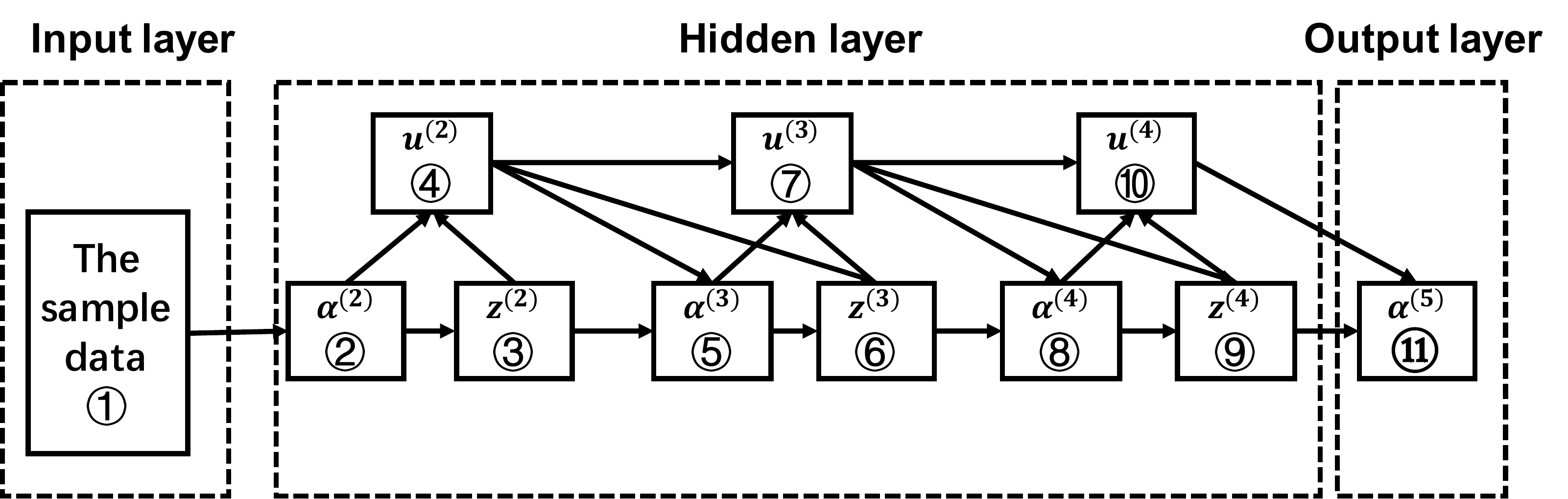}
	\caption{ An example of Adaptive Sparse Deep Network. It mainly consists of three parts, namely: input layer, hidden layer and output layer.}
	\label{net}
\end{figure*}

In Forward Network, each node belongs to different operation.
Each node can receive values from the former of nodes and produce the output values to the following nodes.
There is no feedback in the whole network during training.
The value propagates unidirectionally from the input layer to the output layer, which can be represented by a directed acyclic graph.
As shown in Fig.\ref{net}, it is the Adaptive Sparse Deep Network.
The under-sampled data from HSIs flows over the input layer to output layer in a order from circled number 1 to number 10, followed by a final sparse vector with circled number 11 and generate the optimal sparse vector.
Then we can obtain the classification of HSIs.
Fig.\ref{net} greatly illustrates the deep architecture described above.

After training the Adaptive Sparse Deep Network, there exists errors in the results.
We can minimize losses by using gradient descent method in Back-Propagation.
By computing the loss function, the parameter of each node can be updated.
Therefore, we get the latest parameters to train the network against.
The optimization of the parameters is expected to obtain optimal sparse vector, which produces improved classification result.

Firstly, the each class of residual error between the network output and truth is defined as
\begin{equation}\label{eq:residual}
{r_i} = \frac{1}{2}\left\| {x - {A_i}{\alpha _i}} \right\|_2^2
\end{equation}

\noindent Then the loss function can be defined as:
\begin{equation}\label{eq:loss function}
E(\Theta ) =  - \mathop {\sum {{y_i}} }\limits_i \log \frac{{{e^{ - {r_i}}}}}{{\sum\limits_i {{e^{ - {r_i}}}} }}
\end{equation}

\noindent where $\Theta$ are the network parameters.
In the backward pass, we aim to learn the following parameters: ${\rho ^{(n)}}$ in sparsity operation, ${\eta ^{(n)}}$ in nonlinear transform operation and ${\tau ^{(n)}}$ in multiplier update operation.
In the following subsection, we discuss how to compute the gradients of the loss function $E(\Theta )$ and the parameters $\Theta$ can be updated by using Back-propagation over the data flow graph.

\begin{figure}[!htbp]
	\centering
	\subfigure[]{\includegraphics[width=1.4in,height=1.4in]{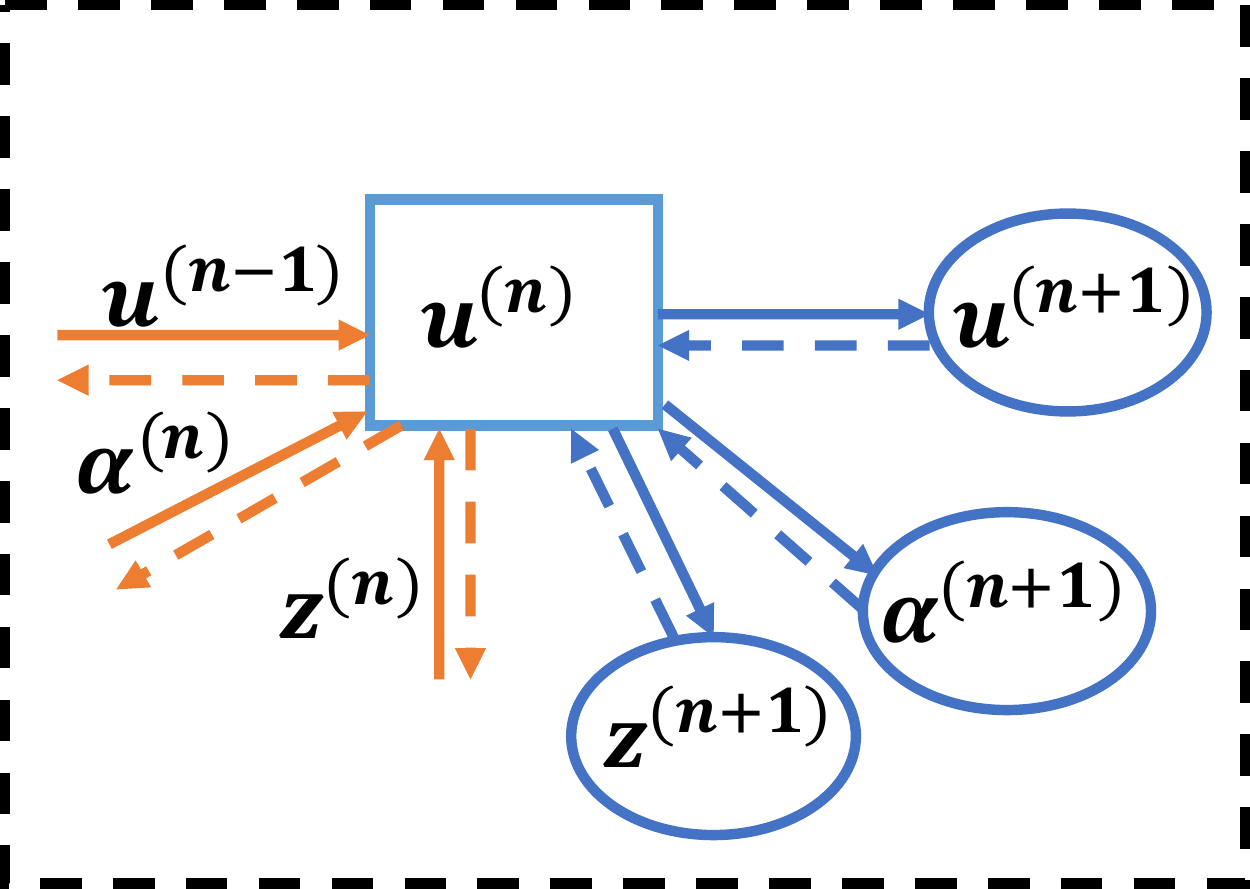}}
	\subfigure[]{\includegraphics[width=1.4in,height=1.4in]{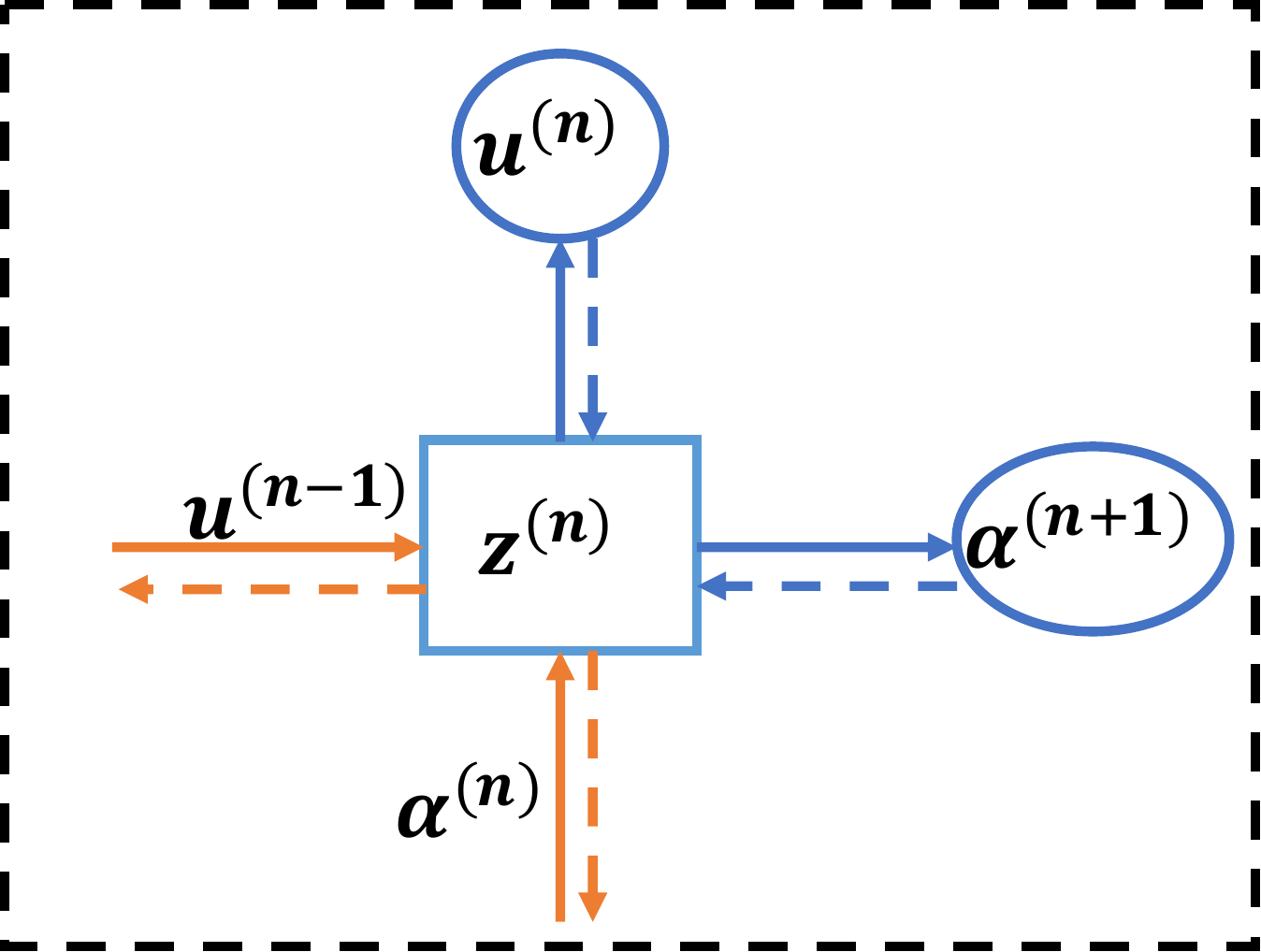}}
	\subfigure[]{\includegraphics[width=1.4in,height=1.4in]{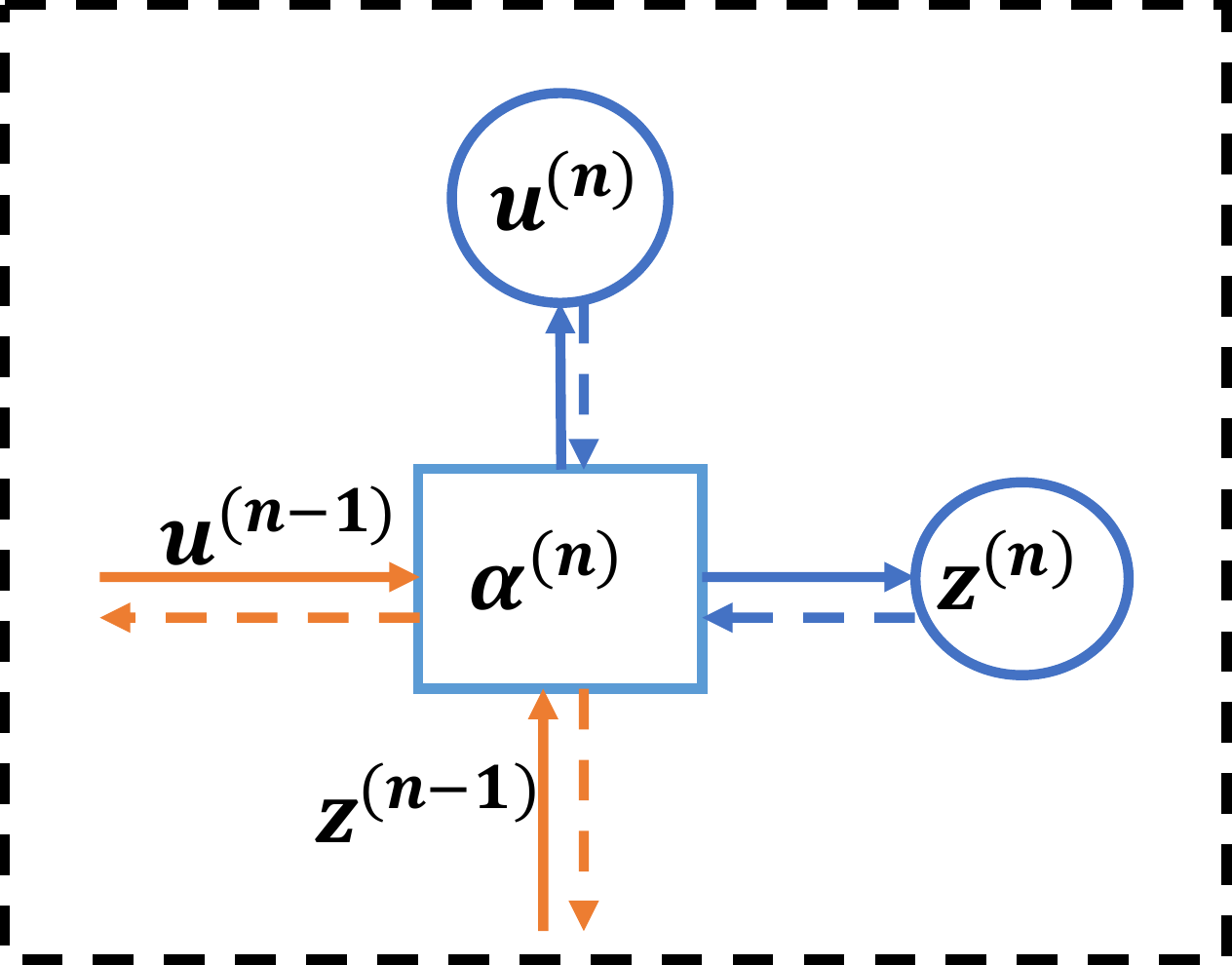}}
	\caption{shows three types of nodes and the data flow over them:(a)Multiplier operation,(b)Nonlinear transform operation and (c)Sparsity operation.}
	\label{loss}
\end{figure}

\subsection{Gradient Computation by Back-Propagation}

In FNN, the gradient propagates backward through each layer, further calculating the gradient of each layer’s parameters.
Similarity, in our Adaptive Sparse Deep Network, the gradients are computed in an inverse order.
Then the parameters can be learned by Back-Propagation.
Fig.\ref{net} shows an example, where the gradient can be computed backwardly from the operation with circled number 11 to 2 successively.
For a stage, Fig.\ref{loss} shows three types of nodes and the data flow over them.
Each node has multiple inputs and outputs.
We next introduce the detail of gradients computation for each node in a typical stage.
%

Multiplier update operation: As shown in Fig.\ref{loss}(a), this operation has three inputs: ${\alpha ^{(n)}}$,${z^{(n)}}$ and ${u^{(n - 1)}}$.
Its output ${u^{(n)}}$ is the input to compute ${\alpha ^{(n + 1)}}$,${z^{(n + 1)}}$ and ${u^{(n + 1)}}$.
Here the parameters can be updated.
The gradients of loss and the parameters can be computed as:
\begin{equation}\label{eq:multiplier updation}
\frac{{\partial E}}{{\partial {\tau ^{(n)}}}} = \frac{{\partial E}}{{\partial {u^{(n)}}}}\frac{{\partial {u^{(n)}}}}{{\partial {\tau ^{(n)}}}}
\end{equation}

\begin{equation}\label{eq:multiplier updation(2)}
{\rm{ }}\frac{{\partial E}}{{\partial {u^{(n)}}}} = \frac{{\partial E}}{{\partial {u^{(n + 1)}}}}\frac{{\partial {u^{(n + 1)}}}}{{\partial {u^{(n)}}}} + \frac{{\partial E}}{{\partial {\alpha ^{(n + 1)}}}}\frac{{\partial {\alpha ^{(n + 1)}}}}{{\partial {u^{(n)}}}}
{\rm{                  }}   + \frac{{\partial E}}{{\partial {z^{(n + 1)}}}}\frac{{\partial {z^{(n + 1)}}}}{{\partial {u^{(n)}}}}
\end{equation}
We also compute gradient of the output in this operation and its inputs:$\frac{{\partial {u^{(n)}}}}{{\partial {u^{(n - 1)}}}}$,$\frac{{\partial {u^{(n)}}}}{{\partial {\alpha ^{(n)}}}}$ and  $\frac{{\partial {u^{(n)}}}}{{\partial {z^{(n)}}}}$.

Nonlinear transform update operation: As shown in Fig.\ref{loss}(b), this operation has two inputs: ${\alpha ^{(n)}}$,${u^{(n - 1)}}$ and its output ${z^{(n)}}$ is the input for computing ${{u^{(n)}}}$ and ${\alpha ^{(n+1)}}$ in next stage.
Here the parameters ${\eta ^{(n)}}$ can be updated.
The gradients of loss and the parameters can be computed as:
\begin{equation}\label{eq:nonlinear updation}
\frac{{\partial E}}{{\partial {\eta ^{(n)}}}} = \frac{{\partial E}}{{\partial {z^{(n)}}}}\frac{{\partial {z^{(n)}}}}{{\partial {\eta ^{(n)}}}}
\end{equation}

\begin{equation}\label{eq:nonlinear updation(2)}
{\rm{ }}\frac{{\partial E}}{{\partial {z^{(n)}}}} = \frac{{\partial E}}{{\partial {u^{(n)}}}}\frac{{\partial {u^{(n)}}}}{{\partial {z^{(n)}}}} + \frac{{\partial E}}{{\partial {\alpha ^{(n + 1)}}}}\frac{{\partial {\alpha ^{(n + 1)}}}}{{\partial {z^{(n)}}}}
\end{equation}
We also compute the gradients of this operation output to its inputs: $\frac{{\partial {z^{(n)}}}}{{\partial {\alpha ^{(n)}}}}$ and $\frac{{\partial {z^{(n)}}}}{{\partial {u^{(n - 1)}}}}$.

Sparsity update operation: As shown in Fig.\ref{loss}(c), this operation has two inputs: ${z^{(n-1)}}$,${{u^{(n-1)}}}$ and its output ${\alpha ^{(n)}}$ is the input for computing  ${z^{(n)}}$ and ${{u^{(n)}}}$ in the same stage.
Here the parameters ${{\rho ^{(n)}}}$ can be updated.
The gradients of loss and the parameters can be computed as:
\begin{equation}\label{eq:sparsity updation}
\frac{{\partial E}}{{\partial {\rho ^{(n)}}}} = \frac{{\partial E}}{{\partial {\alpha ^{(n)}}}}\frac{{\partial {\alpha ^{(n)}}}}{{\partial {\rho ^{(n)}}}}
\end{equation}

\begin{equation}\label{eq:sparsity updation(2)}
{\rm{ }}\frac{{\partial E}}{{\partial {\alpha ^{(n)}}}} = \frac{{\partial E}}{{\partial {z^{(n)}}}}\frac{{\partial {z^{(n)}}}}{{\partial {\alpha ^{(n)}}}} + \frac{{\partial E}}{{\partial {u^{(n)}}}}\frac{{\partial {u^{(n)}}}}{{\partial {\alpha ^{(n)}}}}
\end{equation}
The gradients of this operation output to its inputs:$\frac{{\partial {\alpha ^{(n)}}}}{{\partial {z^{(n - 1)}}}}$ and $\frac{{\partial {\alpha ^{(n)}}}}{{\partial {u^{(n - 1)}}}}$.

\begin{table}[!htbp]
	\caption{Nine classes in the ROSIS Urban Pavia University data and the training and test set for each class}
	\centering
	\begin{tabular}{ccccc}
		\toprule
		Class & Name &Dictionary &Train &Test\\
		\midrule
		1\cellcolor{unicolor1} &Asphalt                  &66        &597  &5968\\
		2\cellcolor{unicolor2} &Meadows                  &186       &932  &17531 \\
		3\cellcolor{unicolor3} &Gravel                   &21        &189  &1889\\
		4\cellcolor{unicolor4} &Trees                    &31        &276  &2757\\
		5\cellcolor{unicolor5} &Painted metal sheets     &13        &269  &1063\\
		6\cellcolor{unicolor6} &Bare Soil                &50        &453  &4526\\
		7\cellcolor{unicolor7} &Bitumen                  &13        &266  &1051\\
		8\cellcolor{unicolor8} &Self-Blocking Bricks     &37        &331  &3314\\
		9\cellcolor{unicolor9} &Shadows                  &9         &189  &749\\
		\midrule
		&Total                    &426        &3502 &38848\\
		\bottomrule
		\label{pavia}
	\end{tabular}
\end{table} 

\begin{figure}[!htbp]
	\centering
	\subfigure[]{\includegraphics[width=0.8in,height=1.6in]{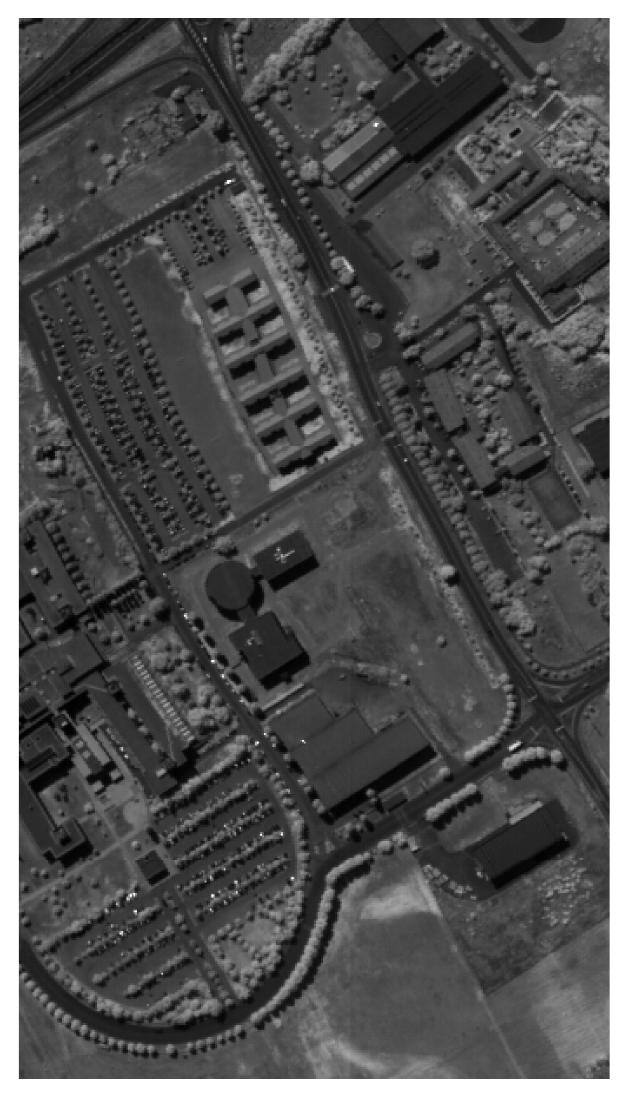}}
	\subfigure[]{\includegraphics[width=0.8in,height=1.6in]{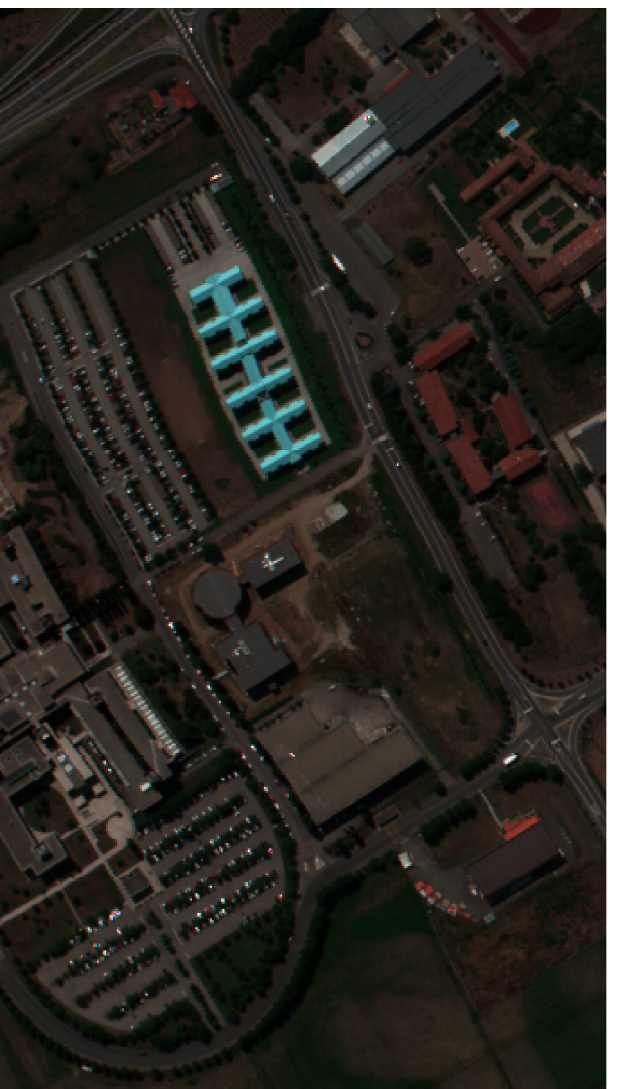}}
	\subfigure[]{\includegraphics[width=0.8in,height=1.6in]{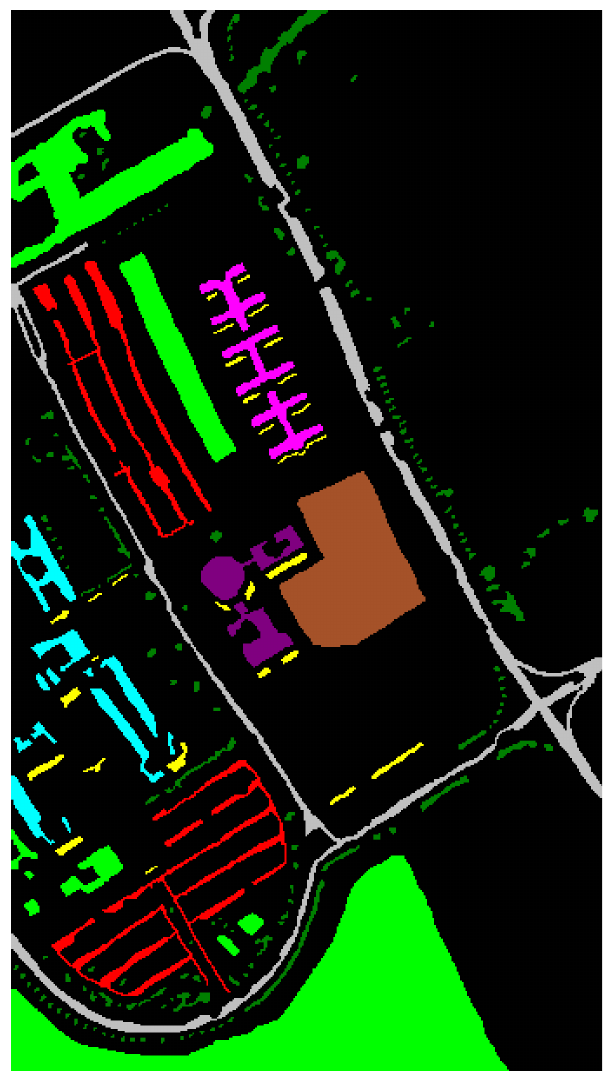}}
	\caption{Reference map of Pavia University. (a)first principal component map of Pavia University; (b)false color image; (c)label map of the ground truth}
	\label{Pavia University}
\end{figure}

\begin{table}[!htbp]
	\caption{Sixteen classes in the AVIRIS Salinas data and the training and test set for each class}
	\centering
	\begin{tabular}{ccccc}
		\toprule
		Class & Name &Dictionary &Train &Test\\
		\midrule
		1\cellcolor{sacolor1}  &Weeds-1                   &20       &181  &1808 \\
		2\cellcolor{sacolor2}  &Weeds-2                   &37       &335  &3334 \\
		3\cellcolor{sacolor3}  &Fallow                    &20       &178  &1778\\
		4\cellcolor{sacolor4}  &Fallow-rough-plow         &14       &125  &1255\\
		5\cellcolor{sacolor5}  &Fallow-smooth             &27       &241  &2410\\
		6\cellcolor{sacolor6}  &Stubble                   &40       &356  &3563\\
		7\cellcolor{sacolor7}  &Celery                    &36       &322  &3221\\
		8\cellcolor{sacolor8}  &Grapes-untrained          &113       &1014 &10144\\
		9\cellcolor{sacolor9}  &Soil-vinyard-develop      &62       &558  &5583\\
		10\cellcolor{sacolor10}&Corn-senesced-green weeds &33       &295  &2950\\
		11\cellcolor{sacolor11}&Lettuce-romaine-4wk       &11       &96  &961\\
		12\cellcolor{sacolor12}&Lettuce-romaine-5wk       &19       &173  &1735\\
		13\cellcolor{sacolor13}&Lettuce-romaine-6wk       &9       &82   &825\\
		14\cellcolor{sacolor14}&Lettuce-romaine-7wk       &11       &96  &963\\
		15\cellcolor{sacolor15}&Vineyard-untrained        &73        &654  &6541\\
		16\cellcolor{sacolor16}&Vineyard-vertical-trellis &18        &163  &1626\\
		\midrule
		&Total                     &543       &4869 &48697\\
		\bottomrule
		\label{salinas}
	\end{tabular}
\end{table}

\begin{figure}[!htbp]
	\centering
	\subfigure[]{\includegraphics[width=0.8in,height=1.6in]{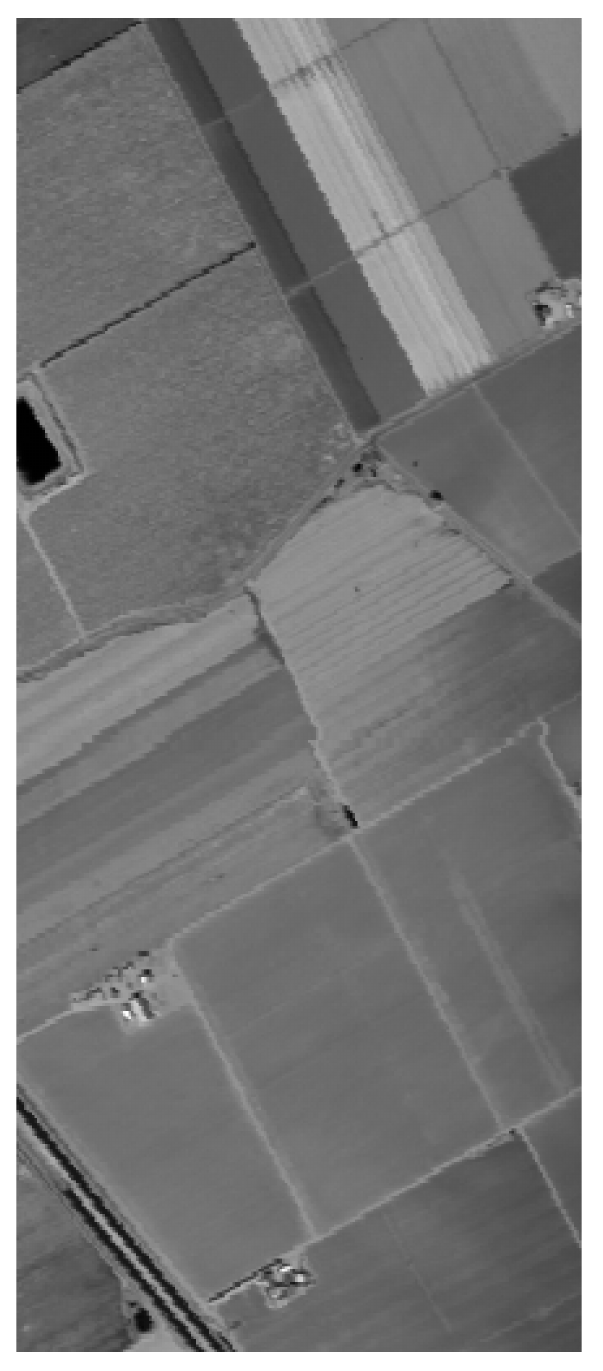}}
	\subfigure[]{\includegraphics[width=0.8in,height=1.6in]{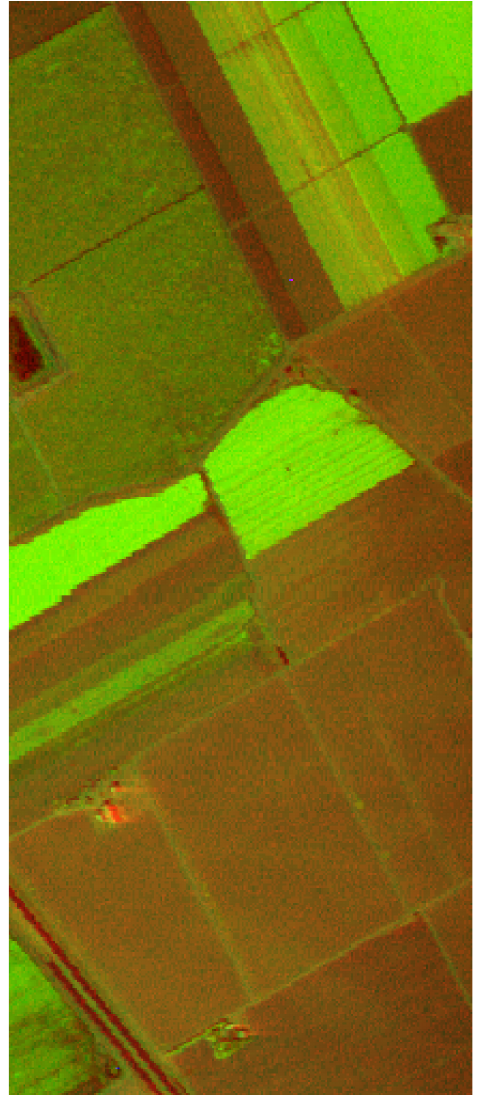}}
	\subfigure[]{\includegraphics[width=0.8in,height=1.6in]{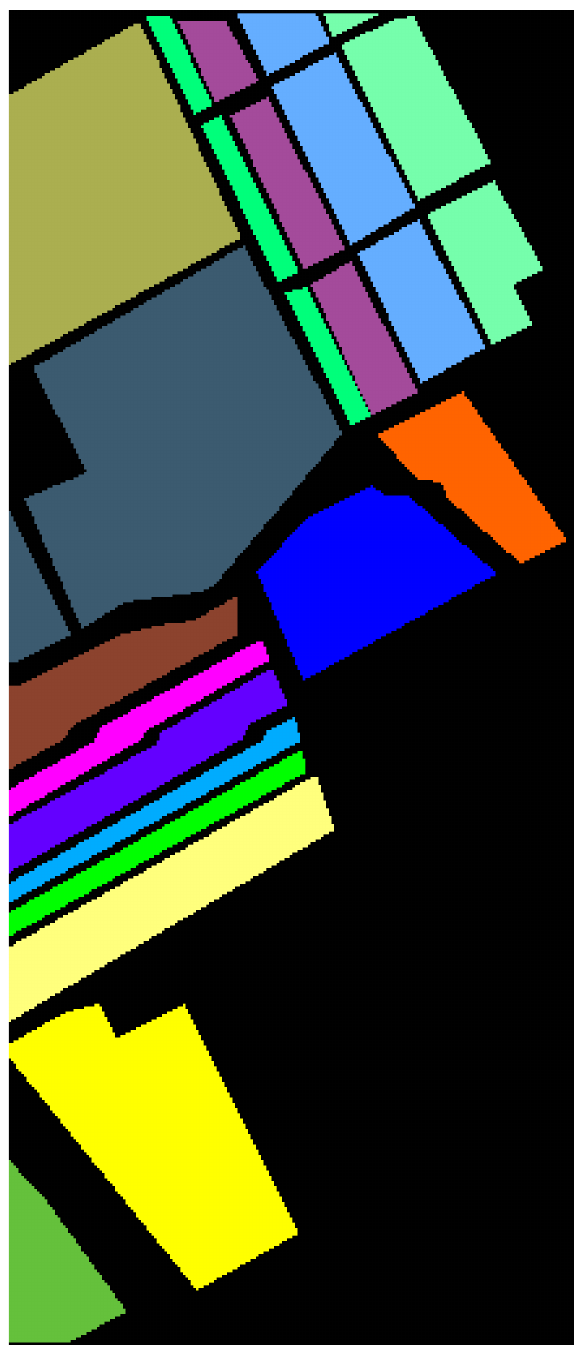}}
	\caption{Reference map of Salinas. (a)first principal component map of Salinas; (b)false color image; (c)label map of the ground truth}
	\label{Salinas}
\end{figure}

\section{ Experimential Results and Analysis}

In this section, four experiments were separately conducted on two real HSIs to validate the superiority of the proposed Adaptive Sparse Deep Network (ASDN).

The first hyperspectral image in our experiments is the University of Pavia, which was captured by the Reflective Optics System Imaging Spectrometer(ROSIS) optical sensor over an urban area surrounding the University of Pavia, Italy.
The size of this image is 610 $\times$ 340 $\times$ 115 with a spatial resolution of 1.3 per pixel and a spectral coverage ranging from 0.43 to 0.86 $\mu m$.
In our experiment, the 12 very noisy channels were removed.
The first principal component map of Pavia University and its false color image are shown in Fig.\ref{Pavia University}(a) and Fig.\ref{Pavia University}(b).
For this date set with 9 classes of land cover, we randomly select 1$\%$ of each class of samples for dictionary and the remainder was used for training and testing.
The reference contents are shown in Table \ref{pavia} and the label map of ground truth is shown in Fig.\ref{Pavia University}(c).

The second hyperspectral image in our experiments is the Salinas, which was acquired by the Airborne/Visible Infrared Imaging Specrometer(AVIRIS) sensor over Salinas Valley, California.
The size of this image is 512 $\times$ 217 $\times$ 224 with a spatial resolution of 3.7 per pixel. Similar to the Pavia University image, 20 water absorption bands(108-112,154-167,and 224) are degraded, and the remaining 204 bands were preserved for the following experiment.
The first principal component map of Pavia University and its false color image are shown in Fig.\ref{Salinas}(a) and Fig.\ref{Salinas}(b).
For this date set with 16 classes of land cover, we randomly select 1$\%$ of each class of samples for dictionary and the remainder was used for training and testing.
The reference contents are shown in Table.\ref{salinas} and the label map of ground truth is shown in Fig.\ref{Salinas}(c).

In these experiments, the influence of different parameters on the classification results will be tested.
The overall accuracy ($OA$), average accuracy ($AA$), and kappa coefficient ($k$) are adopted in these experiments to evaluate the qualify of the classification result.
All the experiments are conducted on a 3.00 GHZ computer with 64.0 Gb RAM.

\subsection{Comparisons of Different Sparsity Levels}

In the first experiment, the proposed ASDN was compared with the greedy algorithms, such as SP~\cite{Dai2009Subspace}, OMP~\cite{Davenport2010Analysis}, ROMP~\cite{Needell2010Signal} and gOMP~\cite{Jian2011Generalized}.
Different sparsity level $K$ were verified to have an impact on the performance of classification results.

\begin{figure*}[!htbp]
	\centering
	\subfigure[]{\includegraphics[width=1.5in,height=1.0in]{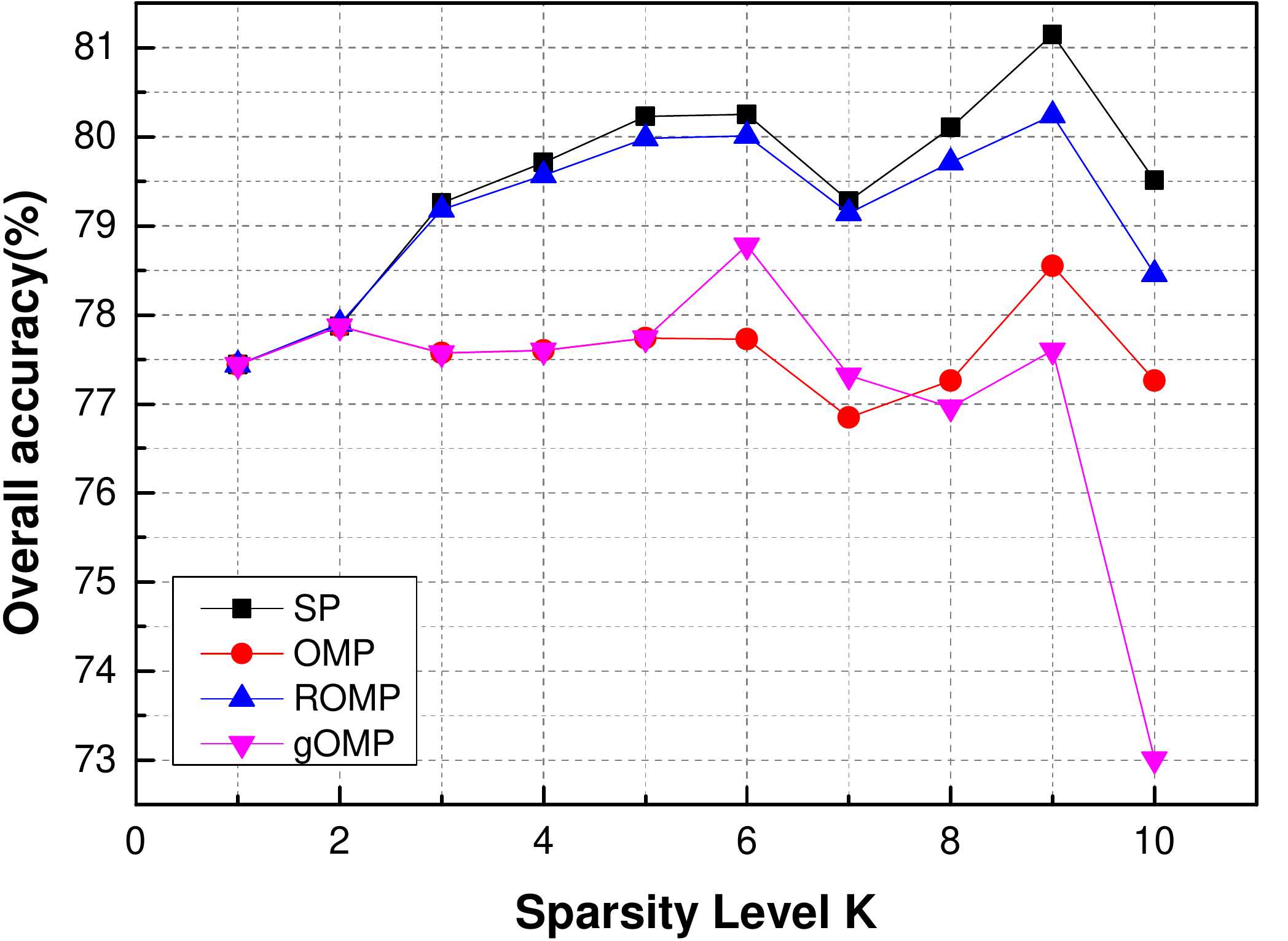}}
	\subfigure[]{\includegraphics[width=1.5in,height=1.0in]{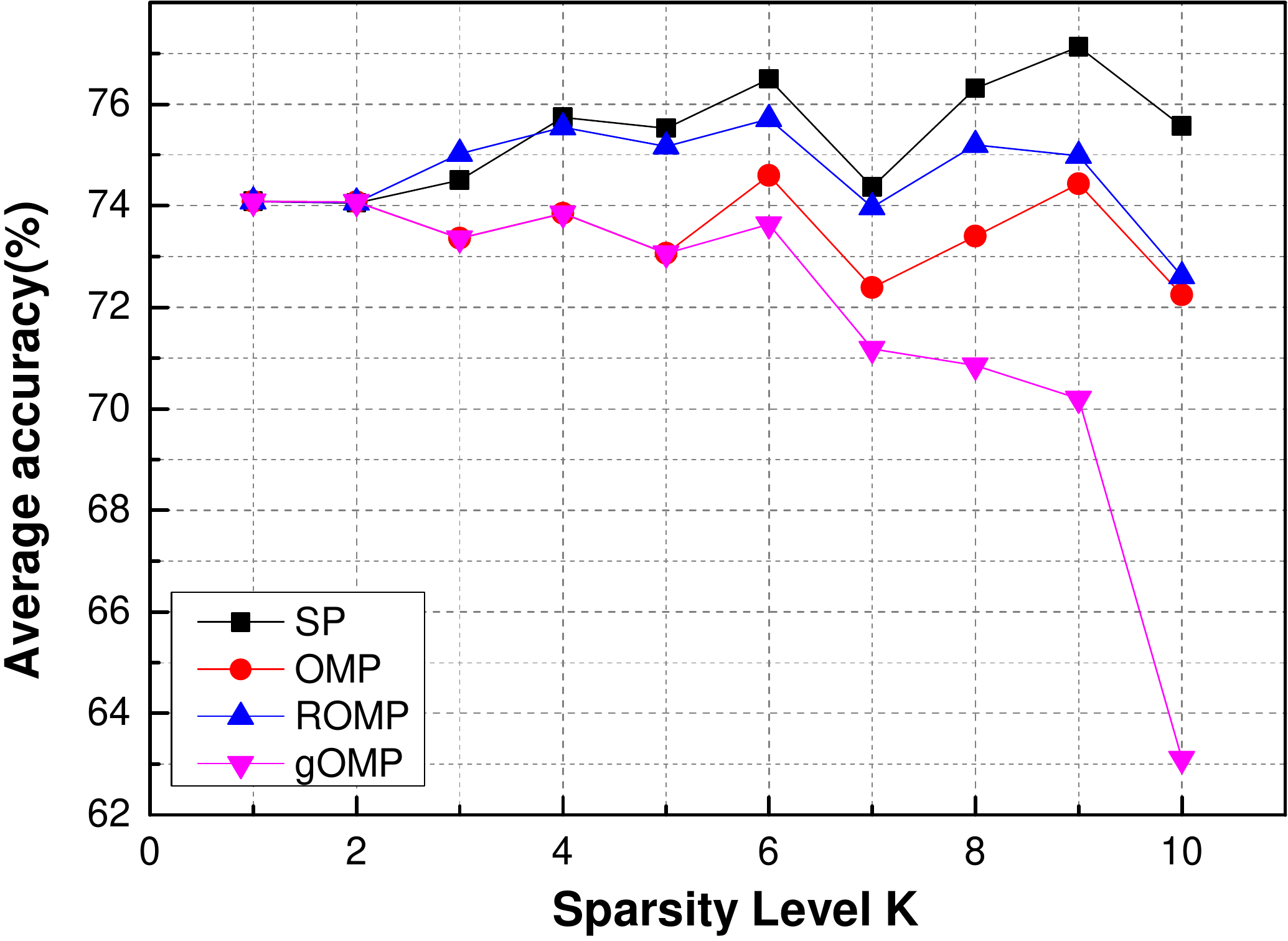}}
	\subfigure[]{\includegraphics[width=1.5in,height=1.0in]{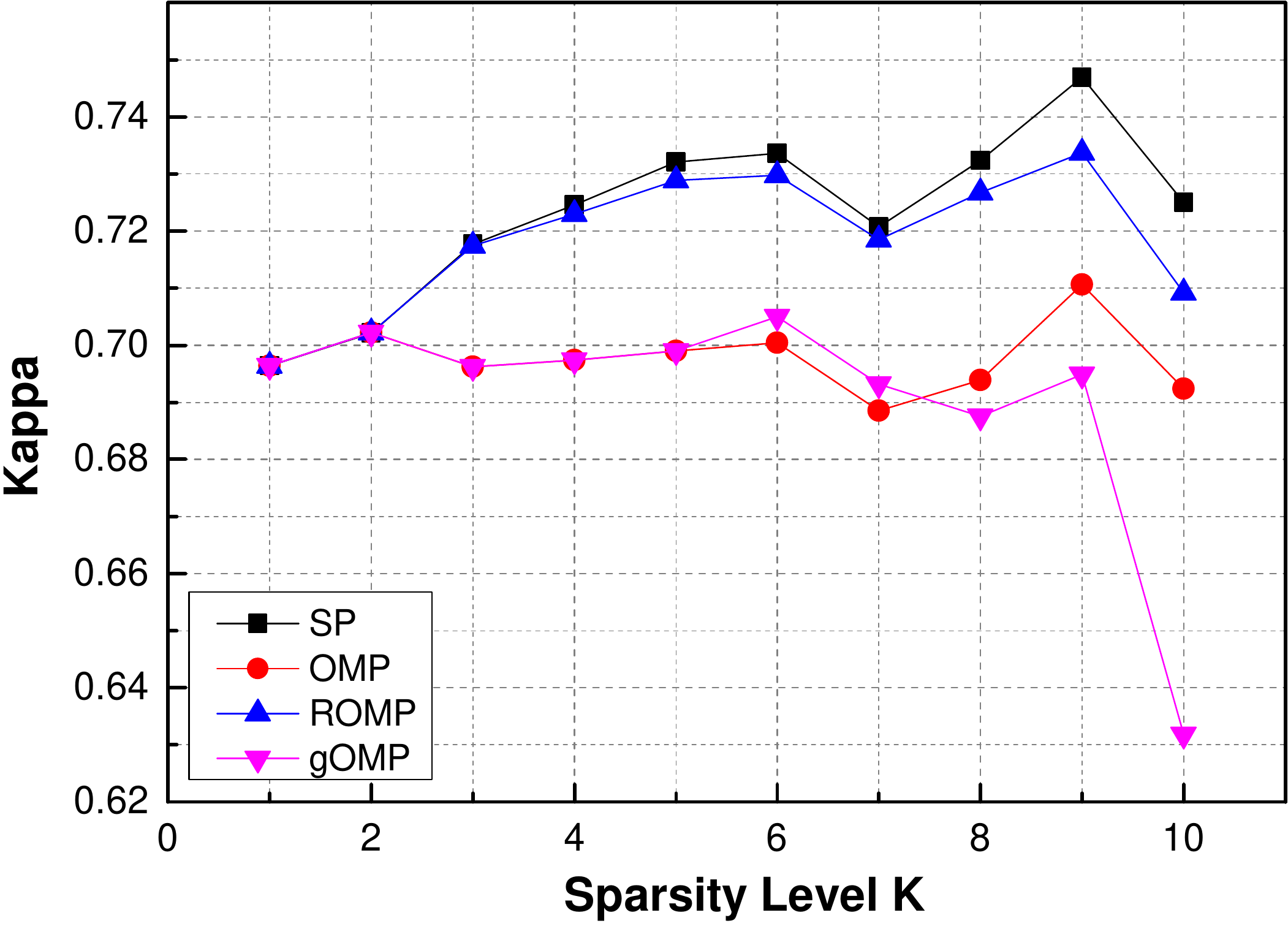}}
	\caption{Different sparsity levels $K$ have influence on the classification results for Pavia University. (a)OA; (b)AA; (c)Kappa}
	\label{Pavia greedy}
\end{figure*}

\begin{figure*}[!htbp]
	\centering
	\subfigure[]{\includegraphics[width=1.5in,height=1.0in]{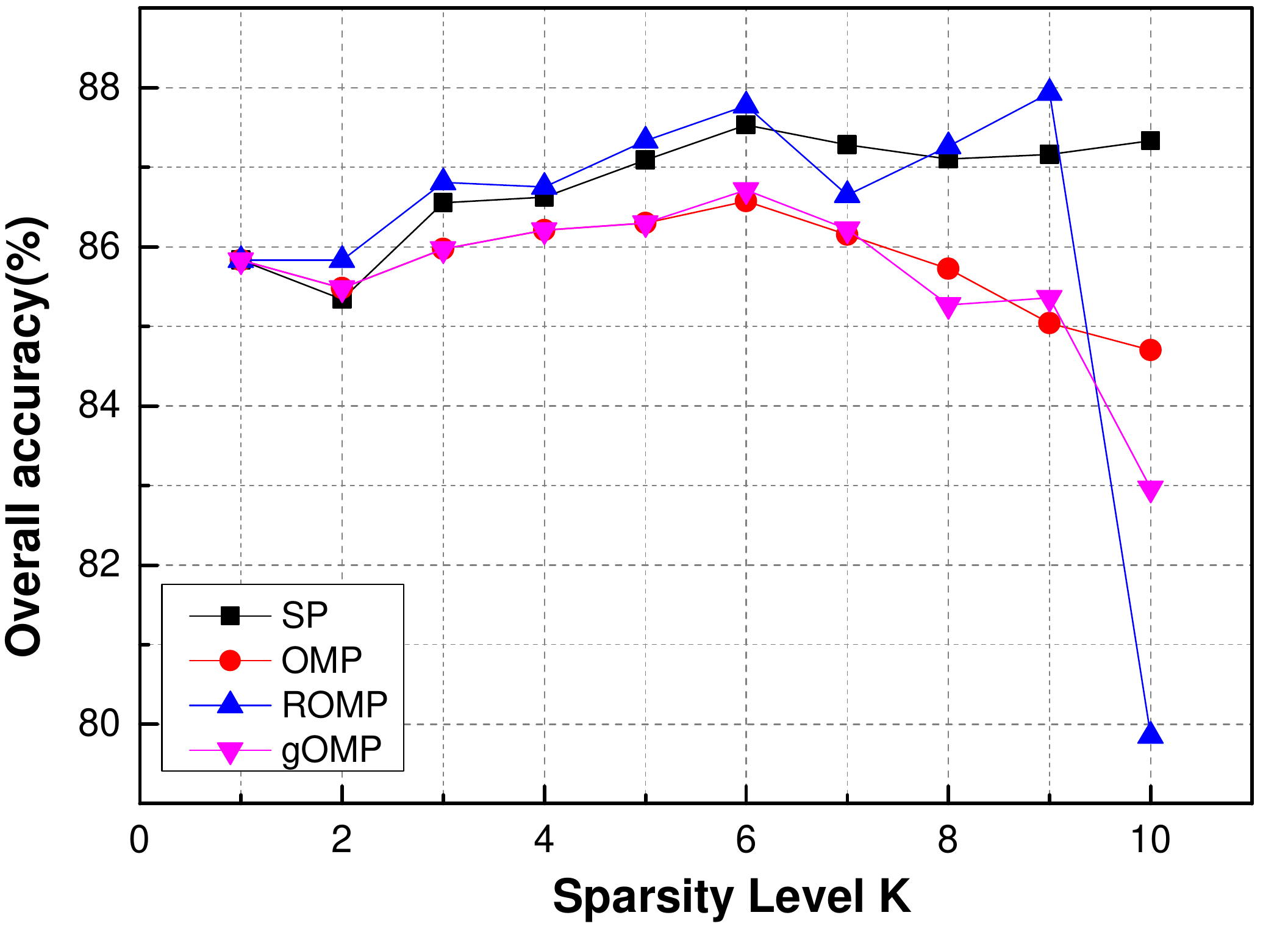}}
	\subfigure[]{\includegraphics[width=1.5in,height=1.0in]{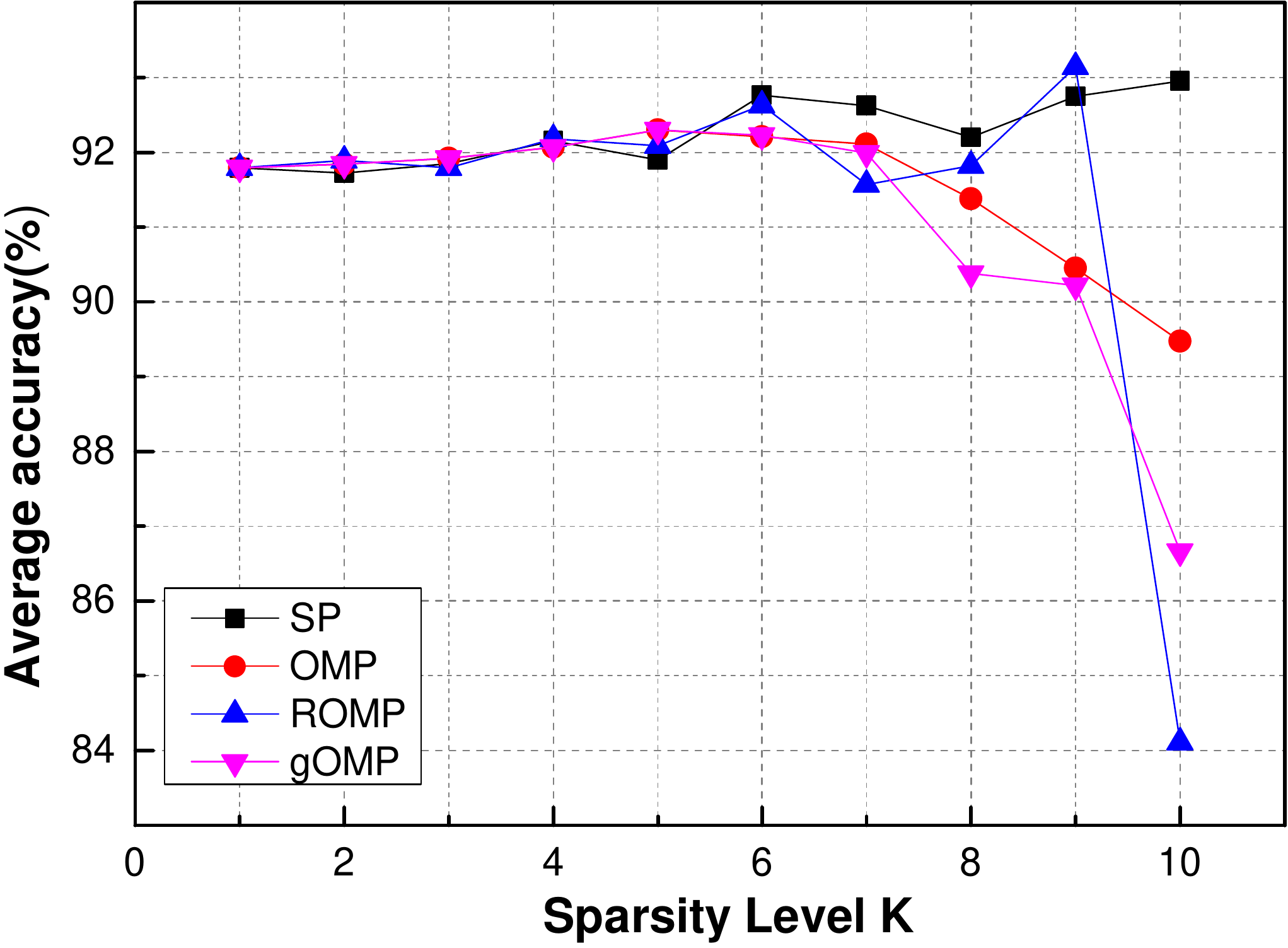}}
	\subfigure[]{\includegraphics[width=1.5in,height=1.0in]{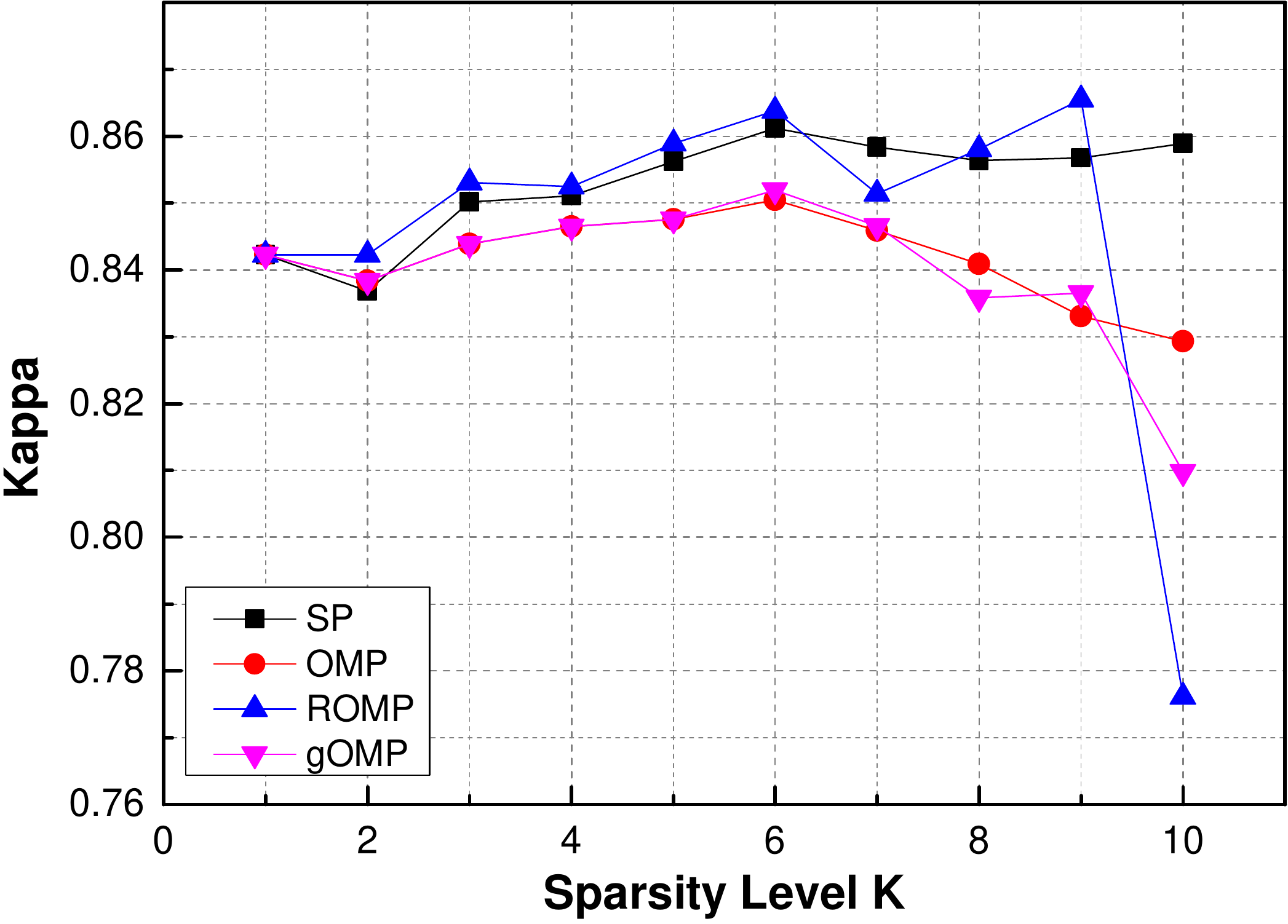}}
	\caption{Different sparsity levels $K$ have influence on the classification results for Salinas. (a)OA; (b)AA; (c)Kappa}
	\label{Salinas greedy}
\end{figure*}

Firstly, we demonstrate the effect of the sparsity level $K$ on the classification results.
Then the results are averaged over five runs at each $K$ to avoid any bias induced by random sampling.
These classification accuracy plots on the entire test sets are shown in Fig.\ref{Pavia greedy} and Fig.\ref{Salinas greedy}.
The sparsiy level $K$ ranges from 1 to 10.
In Fig.\ref{Pavia greedy} and Fig.\ref{Salinas greedy}, the curves are volatile and unstable. 
When the sparsity level $K=9$, the classification accuracy achieves higher on OMP algorithm.
We are still unsure that whether this sparsity level is optimal. 
Other greedy algorithms are similar to above.
Therefore, it is the drawback of manually setting parameters.
Secondly, the classification results of our method and several greedy algorithms are shown in the Table \ref{Pavia result21} and Table \ref{Salinas result21}, including OA, AA and kappa coefficient.
In Pavia University, the OA of ASDN is higher than SP, OMP, ROMP, gOMP, which is 3.87$\%$, 6.47$\%$, 4.78$\%$ and 6.84$\%$ respectively.
Similarly, in Salinas, the OA of ASDN is higher than SP, OMP, ROMP, gOMP, which is 1.13$\%$, 2.09$\%$, 0.73$\%$ and 1.95$\%$ respectively.
Although the SP algorithm and ROMP algorithm showed superior performance respectively, our method outperforms these greedy algorithms.

\begin{table}[!htbp]
	\renewcommand\arraystretch{1.1}
	\centering
	\caption{Classification accuracy ($\%$) for Pavia University using different greedy algorithms}
	\begin{tabular}{|cccccc|}
		\hline
		\multirow{2}{*}{Class}  &\multirow{2}{*}{SP}  &\multirow{2}{*}{OMP}  &\multirow{2}{*}{ROMP}  &\multirow{2}{*}{gOMP}  &Adaptive Sparse \\
		&    &     &      &         &Deep Network       \\
		\hline 
		1	&77.22  &70.05  &69.75  &67.82   &76.35\\ 
		
		2	&92.86  &92.71  &94.55  &93.99   &95.33\\ 
		
		3	&58.81  &54.33  &55.49  &55.39   &65.11\\ 
		
		4	&84.46  &78.21  &83.79  &78.13   &\textbf{92.38}\\ 
		
		5	&98.82  &99.45  &99.17  &\textbf{99.70}   &98.93\\ 
		
		6	&53.46  &44.20  &50.65  &42.57   &\textbf{71.65}\\ 
		
		7	&69.88  &72.04  &69.02  &\textbf{72.44}   &67.25\\ 
		
		8	&73.21  &76.93  &76.50  &72.03   &77.33\\ 
		
		9	&85.45  &82.00  &75.88  &80.58   &68.9\\ 
		\hline 
		$OA$	&81.15  &78.55  &80.24  &78.18   &\textbf{85.02}\\ 
		\hline 
		$AA$	&77.13  &74.43  &75.71  &73.63   &\textbf{79.25}\\ 
		\hline 
		$k$	&74.69  &71.06  &73.37  &70.50   &\textbf{80.07}\\ 
		\hline 
	\end{tabular} 
	\label{Pavia result21}
\end{table}%

\begin{table}[!htbp]
	\renewcommand\arraystretch{1.1}
	\centering
	\caption{Classification accuracy ($\%$) for Salinas using different greedy algorithms}
	\begin{tabular}{|cccccc|}
		\hline
		\multirow{2}{*}{Class}  &\multirow{2}{*}{SP}  &\multirow{2}{*}{OMP}  &\multirow{2}{*}{ROMP}  &\multirow{2}{*}{gOMP}  &Adaptive Sparse \\
		&    &     &      &         &Deep Network       \\
		\hline 
		1	&97.57  &97.18  &98.63  &97.45   &\textbf{99.23}\\ 
		
		2	&98.46  &98.32  &\textbf{98.94}  &98.54   &97.58\\ 
		
		3	&93.46  &\textbf{94.51}  &95.45  &93.51   &91.51\\ 
		
		4	&98.72  &97.80  &\textbf{99.11}  &98.02   &97.69\\ 
		
		5	&97.07  &93.84  &95.91  &94.51   &\textbf{97.26}\\ 
		
		6	&99.51  &\textbf{99.80}  &99.72  &99.83   &98.54\\ 
		
		7	&99.25  &99.34  &99.33  &\textbf{99.51}   &99.16\\ 
		
		8	&75.56  &73.82  &78.02  &74.93   &\textbf{85.76}\\ 
		
		9	&98.00  &97.88  &98.73  &98.14   &\textbf{99.34}\\
		
		10	&86.78  &85.57  &\textbf{89.52}  &86.59   &79.80\\ 
		
		11	&95.58  &\textbf{96.37}&93.50  &97.48   &85.85\\ 
		
		12	&\textbf{99.97}  &99.63  &99.11  &99.53   &98.62\\
		
		13	&97.98  &97.39  &\textbf{98.45}  &97.57   &96.72\\
		
		14	&92.04  &90.68  &93.17  &87.35   &\textbf{96.05}\\
		
		15	&\textbf{61.95}  &58.93  &58.06  &57.31   &60.07\\
		
		16	&92.28  &94.31  &94.61  &\textbf{95.36}   &88.31\\     
		\hline 
		$OA$	&87.53  &86.57  &87.93  &86.71   &\textbf{88.66}\\ 
		\hline 
		$AA$	&92.76  &92.21  &\textbf{93.14}  &92.23   &91.97\\ 
		\hline 
		$k$	&86.12  &85.05  &86.55  &85.19   &\textbf{87.34}\\ 
		\hline 
	\end{tabular} 
	\label{Salinas result21}
\end{table}%

\subsection{Comparisons of Different Regularization Parameters}

In the second experiment, the proposed ASDN was compared with the Bregman~\cite{Ye2011Split}, FISTA~\cite{Beck2009A}, and the SpaRSR~\cite{Wright2009Sparse}.
These $\ell_{1}$ algorithms are used to solve our problem \ref{eq:l1} but the corresponding regularization parameters are need to manual setting in advance.
Whether different regularization parameters affect the classification results were tested.

\begin{figure*}[!htbp]
	\centering
	\subfigure[]{\includegraphics[width=1.5in,height=1.0in]{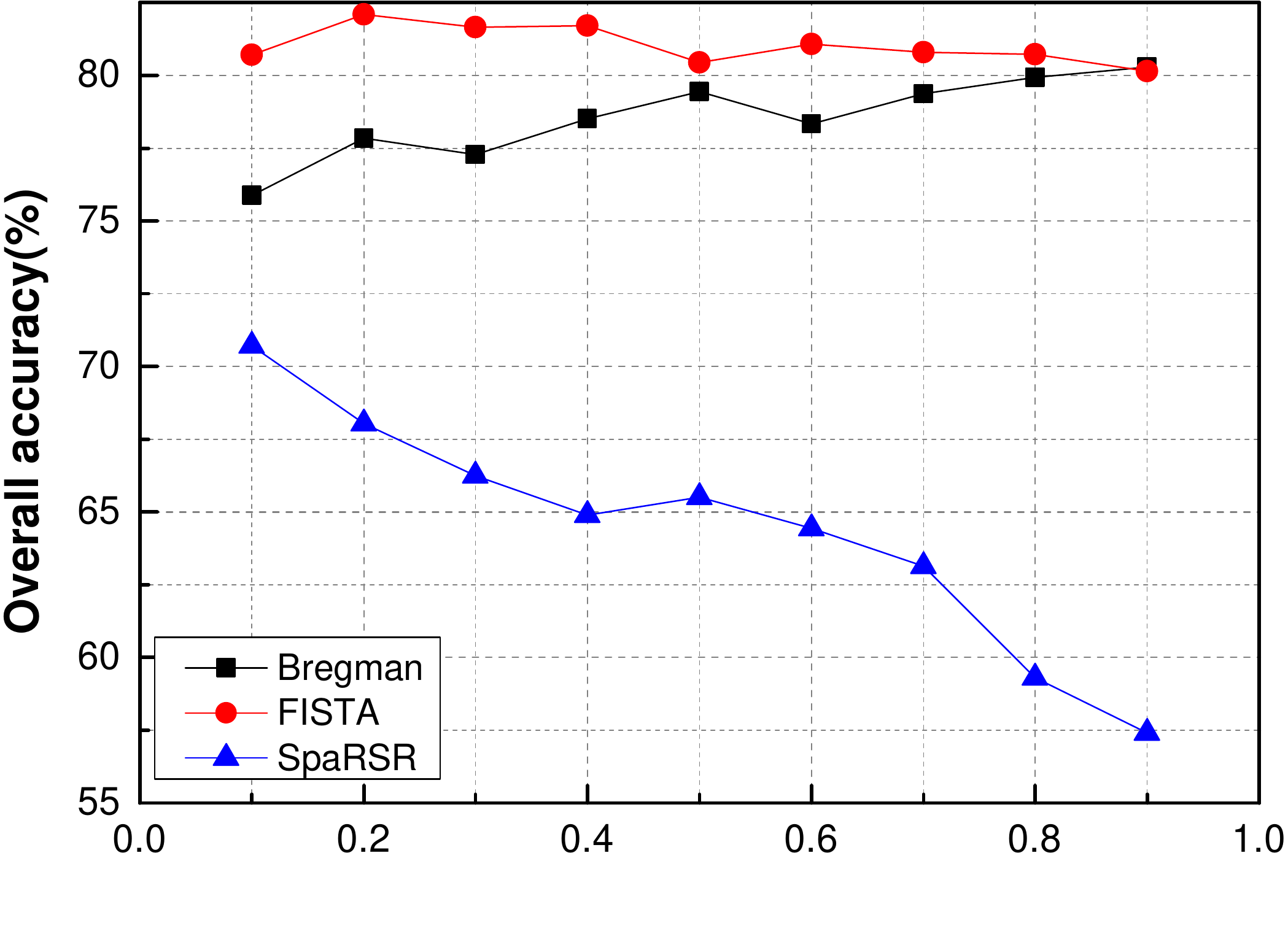}}
	\subfigure[]{\includegraphics[width=1.5in,height=1.0in]{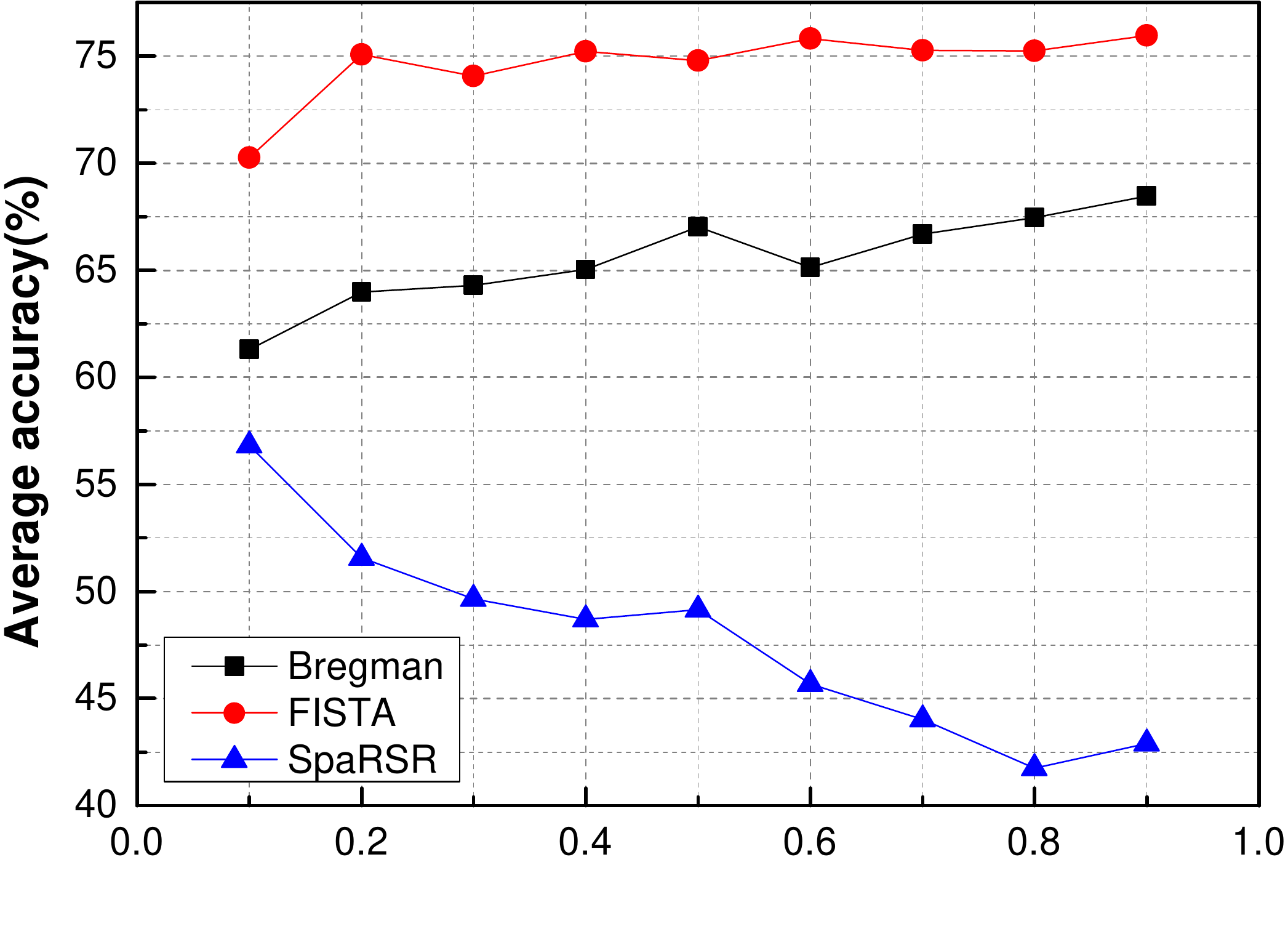}}
	\subfigure[]{\includegraphics[width=1.5in,height=1.0in]{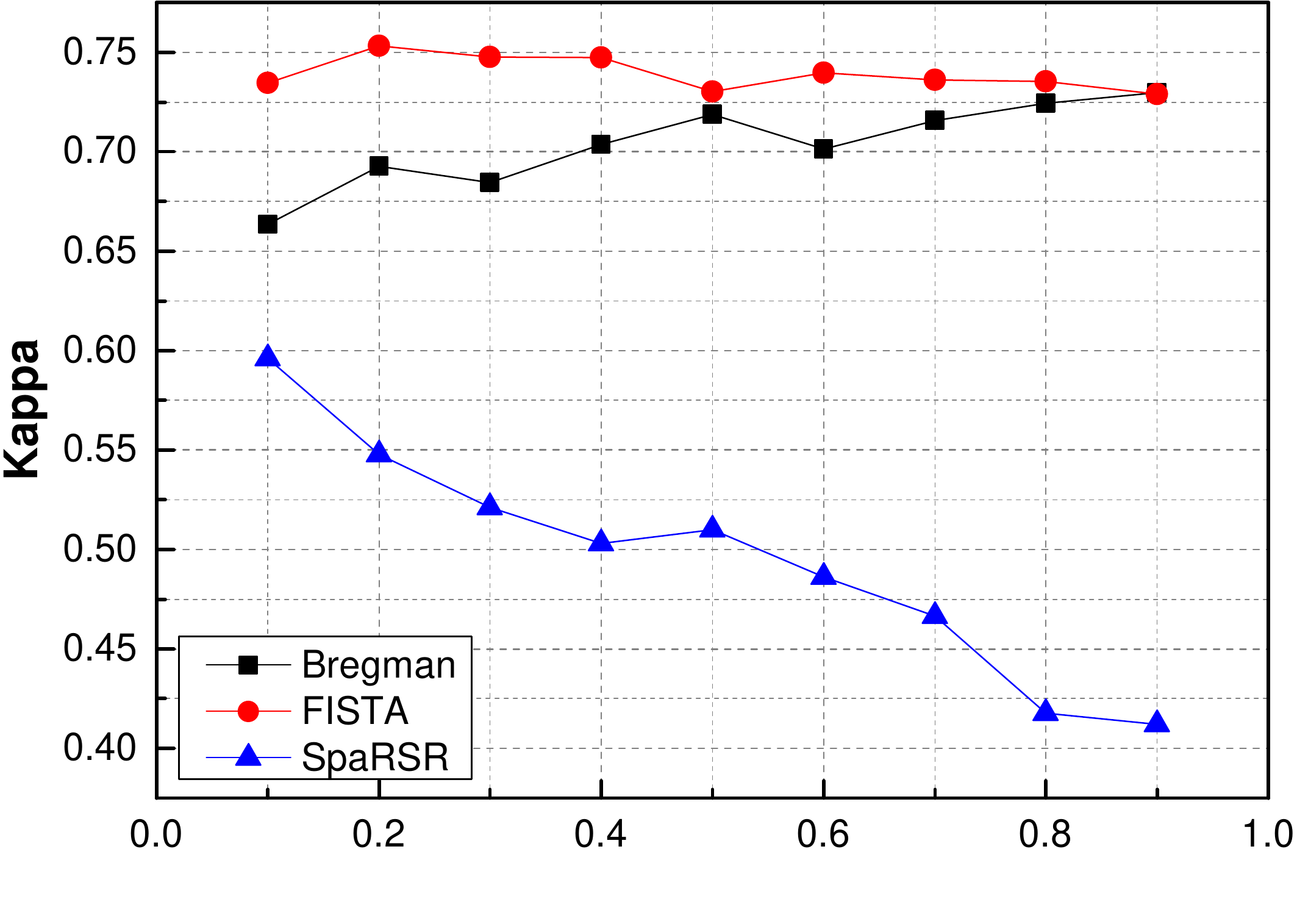}}
	\caption{Different regularization parameters $\lambda$ have influence on the classification results for Pavia University. (a)OA; (b)AA; (c)Kappa}
	\label{Pavia l1}
\end{figure*}

\begin{figure*}[!htbp]
	\centering
	\subfigure[]{\includegraphics[width=1.5in,height=1.0in]{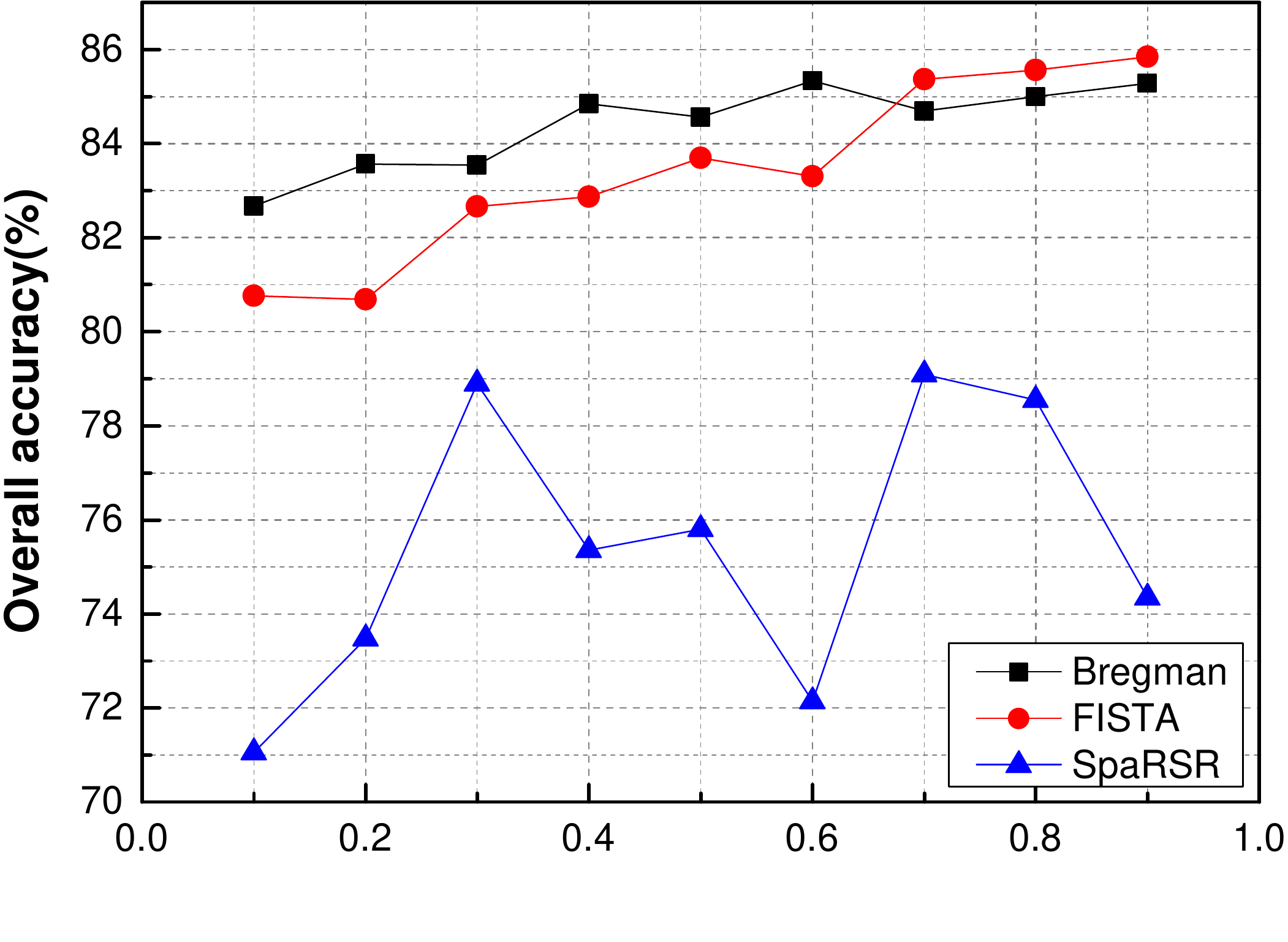}}
	\subfigure[]{\includegraphics[width=1.5in,height=1.0in]{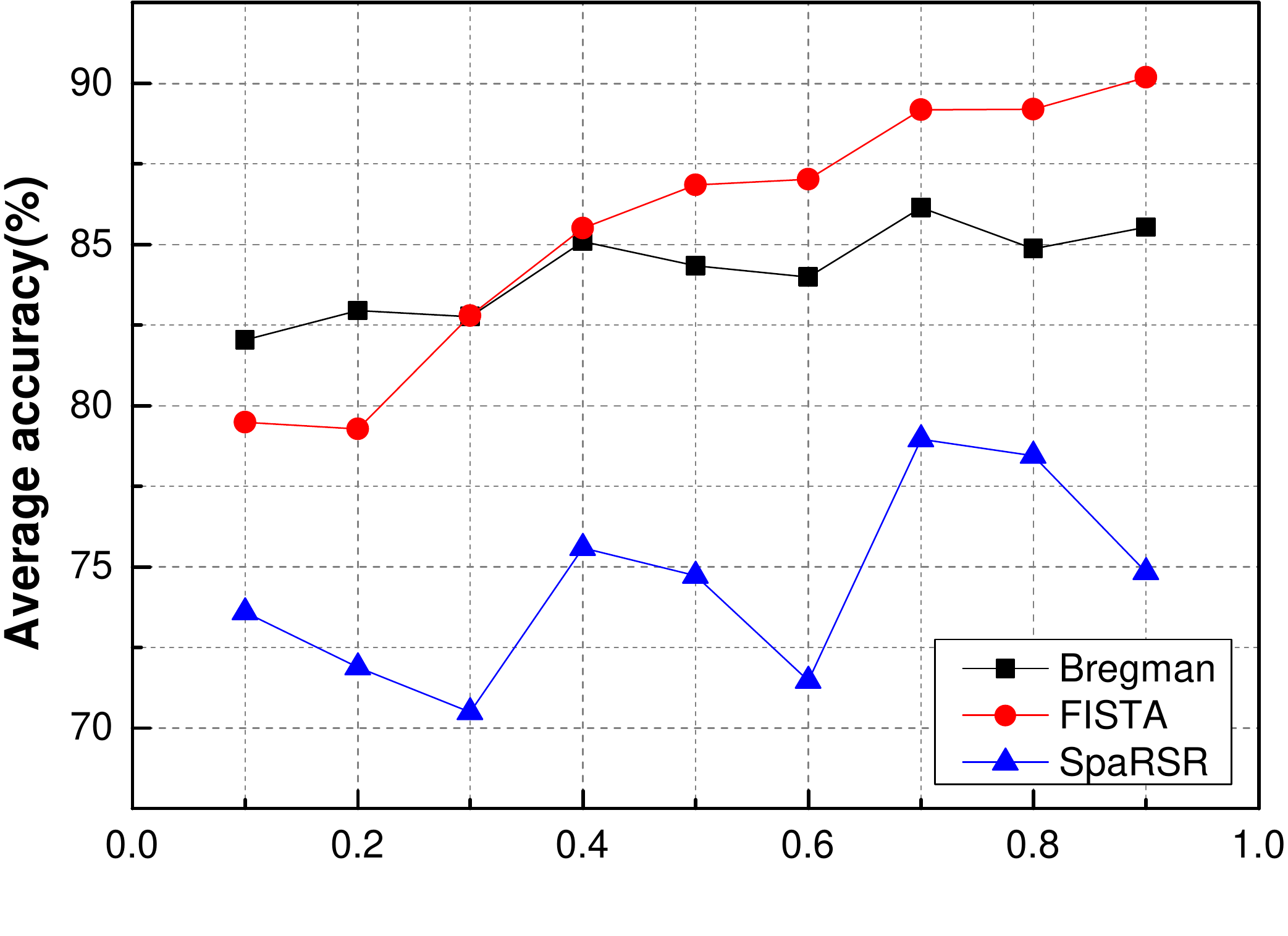}}
	\subfigure[]{\includegraphics[width=1.5in,height=1.0in]{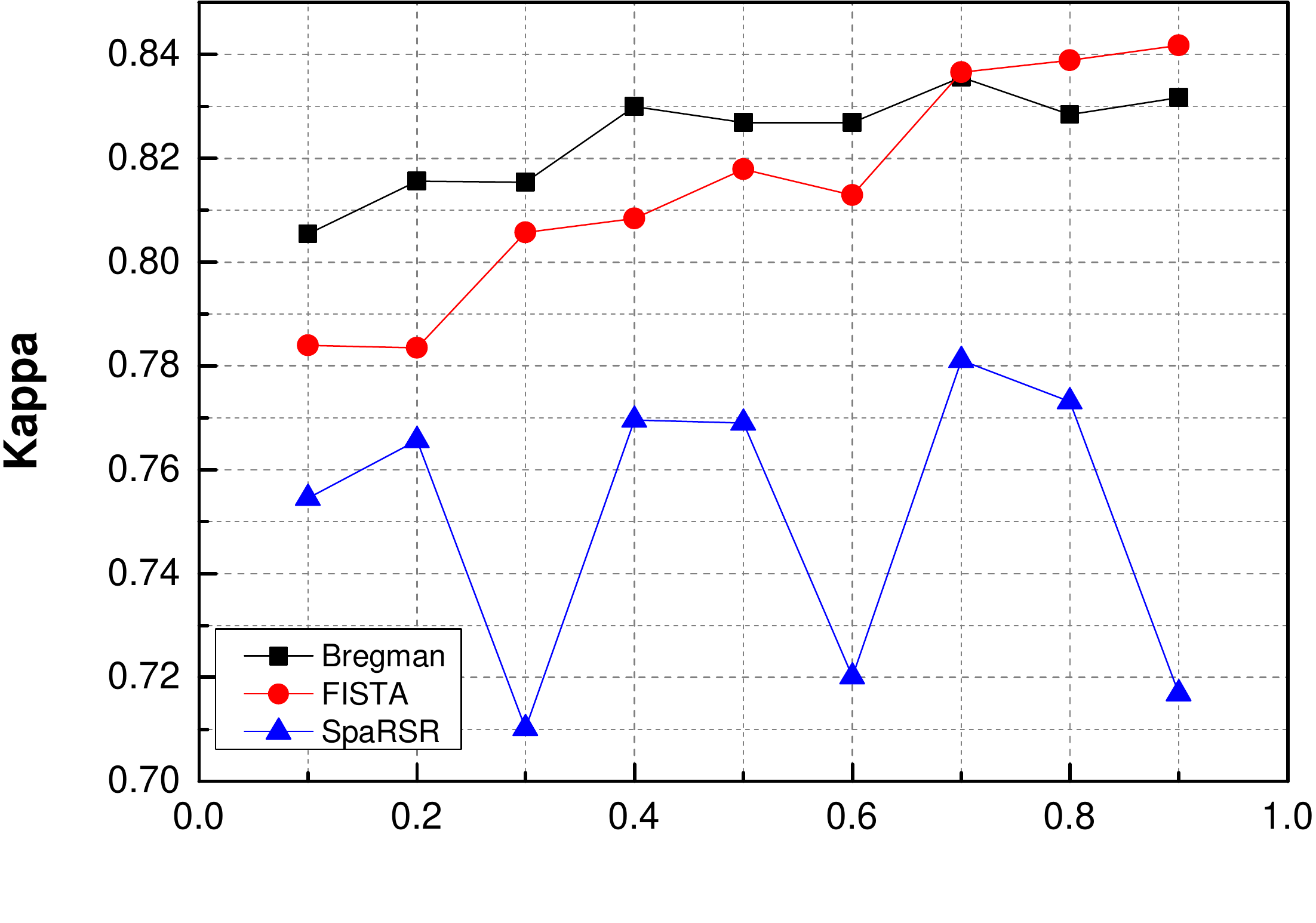}}
	\caption{Different regularization parameters $\lambda$ have influence on the classification results for Salinas. (a)OA; (b)AA; (c)Kappa}
	\label{Salinas l1}
\end{figure*}

Different regularization parameters $\lambda$ are applied to the problem \ref{eq:l1}.
All results are averaged over five runs at each $\lambda$ to avoid any bias induced by random sampling.
The classification plots are shown in Fig.\ref{Pavia l1} and Fig.\ref{Salinas l1}.
As the $\lambda$ increases, the classification curves of the Bregman and FISTA Steadily rise.
But it is hard to determine the optimal regularization parameters.
As for the curves of the SpaRSR fluctuates.
We are still unsure the global optimal solutions.
For these $\ell_{1}$ algorithms, setting the regularization parameters in advance is inevitable.
Then we list the classification results of the proposed ASDN and several $\ell_{1}$ algorithms in Table \ref{Pavia result31} and Table \ref{Salinas result31}.
In terms of OA, the FISTA achieves the highest accuracy except for the proposed ASDN.
Our method is higher than FISTA, which is 2.93$\%$ and 2.81$\%$ in Pavia University and Salinas.
The proposed ASDN still achieves the highest accuracy compared with $\ell_{1}$ algorithms.

\begin{table}[!htbp]
	\renewcommand\arraystretch{1.1}
	\centering
	\caption{Classification accuracy ($\%$) for Pavia University using different  $\ell_{1}$ algorithms and ADMM algorithm}
	\begin{tabular}{|cccccc|}
		\hline
		\multirow{2}{*}{Class}   &\multirow{2}{*}{Bregman} &\multirow{2}{*}{FISTA} &\multirow{2}{*}{SpaRSR} &\multirow{2}{*}{ADMM} &Adaptive Sparse \\
		&        &      &       &    &Deep Network       \\
		\hline 
		1	  &\textbf{94.87} &84.41 &94.20 &88.92 &76.35\\ 
		
		2	  &96.90  &\textbf{98.64} &91.33 &92.13 &95.33\\ 
		
		3	  &\textbf{68.19}  &47.93 &3.01 &48.78 &65.11\\ 
		
		4	  &89.75  &78.55 &80.22 &81.93 &\textbf{92.38}\\ 
		
		5	  &99.65  &99.40 &96.96 &99.40 &98.93\\ 
		
		6	  &47.03  &36.28 &28.23 &52.08 &\textbf{71.65}\\ 
		
		7	  &17.78  &59.35 &0.23 &23.38 &67.25\\ 
		
		8	  &32.69  &\textbf{78.09} &22.11 &45.30 &77.33\\ 
		
		9	  &69.33  &93.12 &\textbf{94.98} &91.71 &68.9\\ 
		\hline 
		$OA$	  &80.29  &82.09 &70.71 &78.12 &\textbf{85.02}\\ 
		\hline 
		$AA$	  &68.47  &75.09 &56.82 &69.29 &\textbf{79.25}\\ 
		\hline 
		$k$	  &72.98  &75.34 &59.60 &70.46 &\textbf{80.07}\\ 
		\hline 
	\end{tabular} 
	\label{Pavia result31}
\end{table}%

\begin{table}[!htbp]
	\renewcommand\arraystretch{1.1}
	\centering
	\caption{Classification accuracy ($\%$) for Salinas using different $\ell_{1}$ algorithms and ADMM algorithm}
	\begin{tabular}{|cccccc|}
		\hline
		\multirow{2}{*}{Class}   &\multirow{2}{*}{Bregman} &\multirow{2}{*}{FISTA} &\multirow{2}{*}{SpaRSR} &\multirow{2}{*}{ADMM} &Adaptive Sparse \\
		&        &      &       &    &Deep Network       \\
		\hline 
		1	  &99.77 &97.48 &79.38 &\textbf{100.00} &99.23\\ 
		
		2	  &95.97  &98.48 &\textbf{99.32} &18.55 &97.58\\ 
		
		3	  &42.0  &79.55 &0.1 &9.30 &\textbf{91.51}\\ 
		
		4	  &98.99  &98.55 &95.07 &\textbf{99.06} &97.69\\ 
		
		5	  &98.87  &97.32 &28.06 &\textbf{99.59} &97.26\\ 
		
		6	  &99.78  &99.29 &99.16 &\textbf{99.64} &98.54\\ 
		
		7	  &\textbf{99.77}  &99.44 &16.48 &99.07 &99.16\\ 
		
		8	  &91.98  &83.25 &\textbf{96.50} &85.98 &85.76\\ 
		
		9	  &99.64  &97.28 &\textbf{99.69} &98.63 &99.34\\ 
		
		10	&\textbf{92.51}  &76.52  &33.65  &45.33   &79.80\\ 
		
		11	&34.41  &\textbf{90.54}  &0  &0   &85.85\\
		
		12	&98.29  &\textbf{99.69}  &8.91  &30.78   &98.62\\
		
		13	&\textbf{99.26}  &98.01  &50.00  &98.90   &96.72\\
		
		14	&92.19  &90.46  &81.11  &88.20   &\textbf{96.05}\\
		
		15	&37.68  &48.41  &4.27  &46.41   &\textbf{60.07}\\
		
		16	&\textbf{97.33}  &88.81  &65.72  &82.49   &88.31\\      
		\hline 
		$OA$	  &85.34  &85.85 &61.05 &72.09 &\textbf{88.66}\\ 
		\hline 
		$AA$	  &86.15  &90.19 &53.59 &68.87 &\textbf{91.97}\\ 
		\hline 
		$k$	  &83.56  &84.18 &55.45 &68.71 &\textbf{87.34}\\ 
		\hline 
	\end{tabular} 
	\label{Salinas result31}
\end{table}%

\subsection{Comparisons of ADMM}

In the third experiment, the proposed ASDN was compared with ADMM on different penalty parameters or regularization parameters.
Firstly, we demonstrated the effect of the penalty parameter $\rho$ on the classification results.
Similarly, whether different regularization parameters $\lambda$ have the impact on the results were tested.

\begin{figure*}[!htbp]
	\centering
	\subfigure[]{\includegraphics[width=2.0in,height=1.5in]{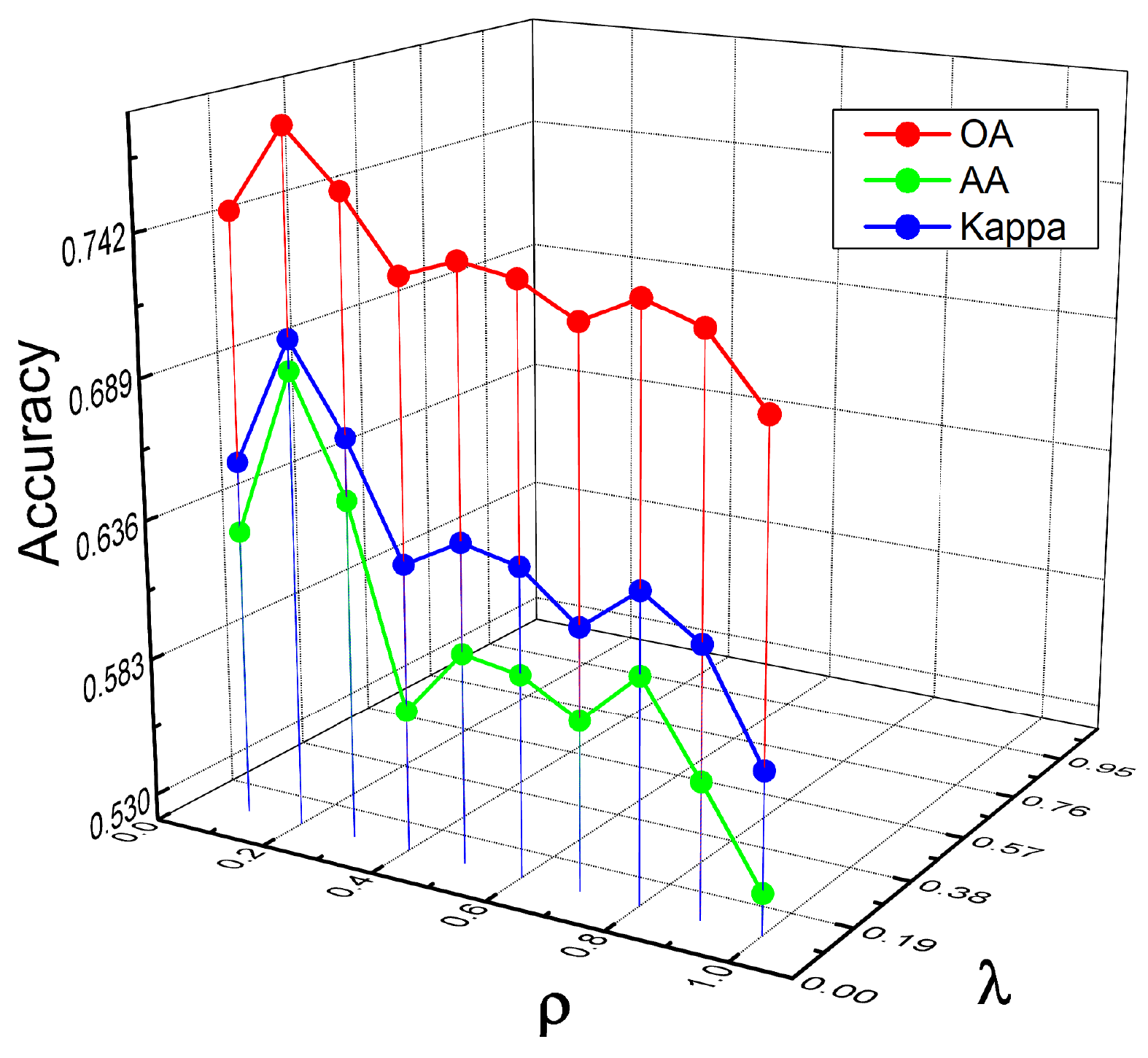}}
	\subfigure[]{\includegraphics[width=2.0in,height=1.5in]{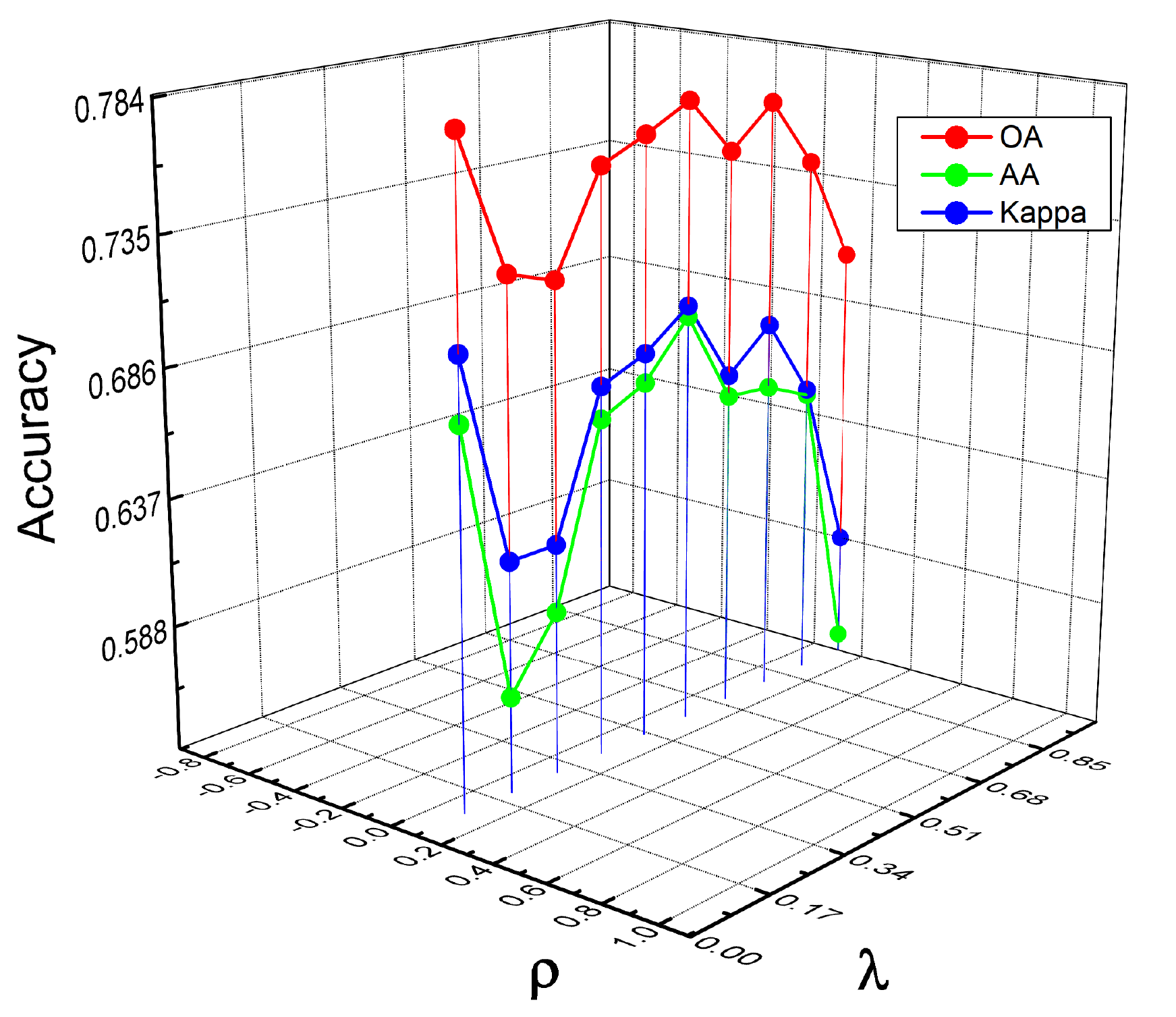}}
	\caption{Classification accuracy for Pavia University using ADMM algorithm. (a)ADMM on different penalty parameters $\rho$; (b)ADMM on different regularization parameters $\lambda$}
	\label{Pavia admm}
\end{figure*}

\begin{figure*}[!htbp]
	\centering
	\subfigure[]{\includegraphics[width=2.0in,height=1.5in]{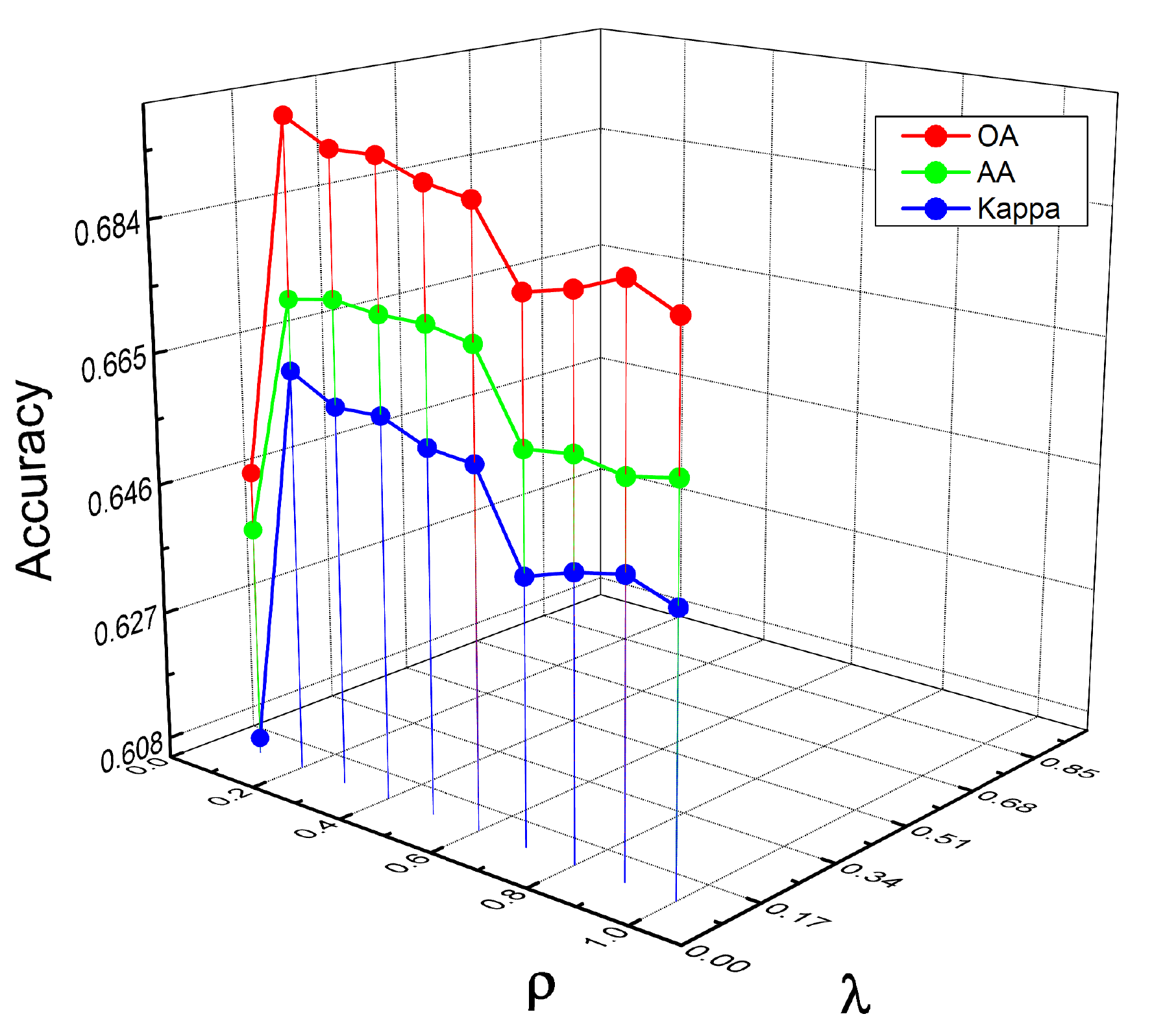}}
	\subfigure[]{\includegraphics[width=2.0in,height=1.5in]{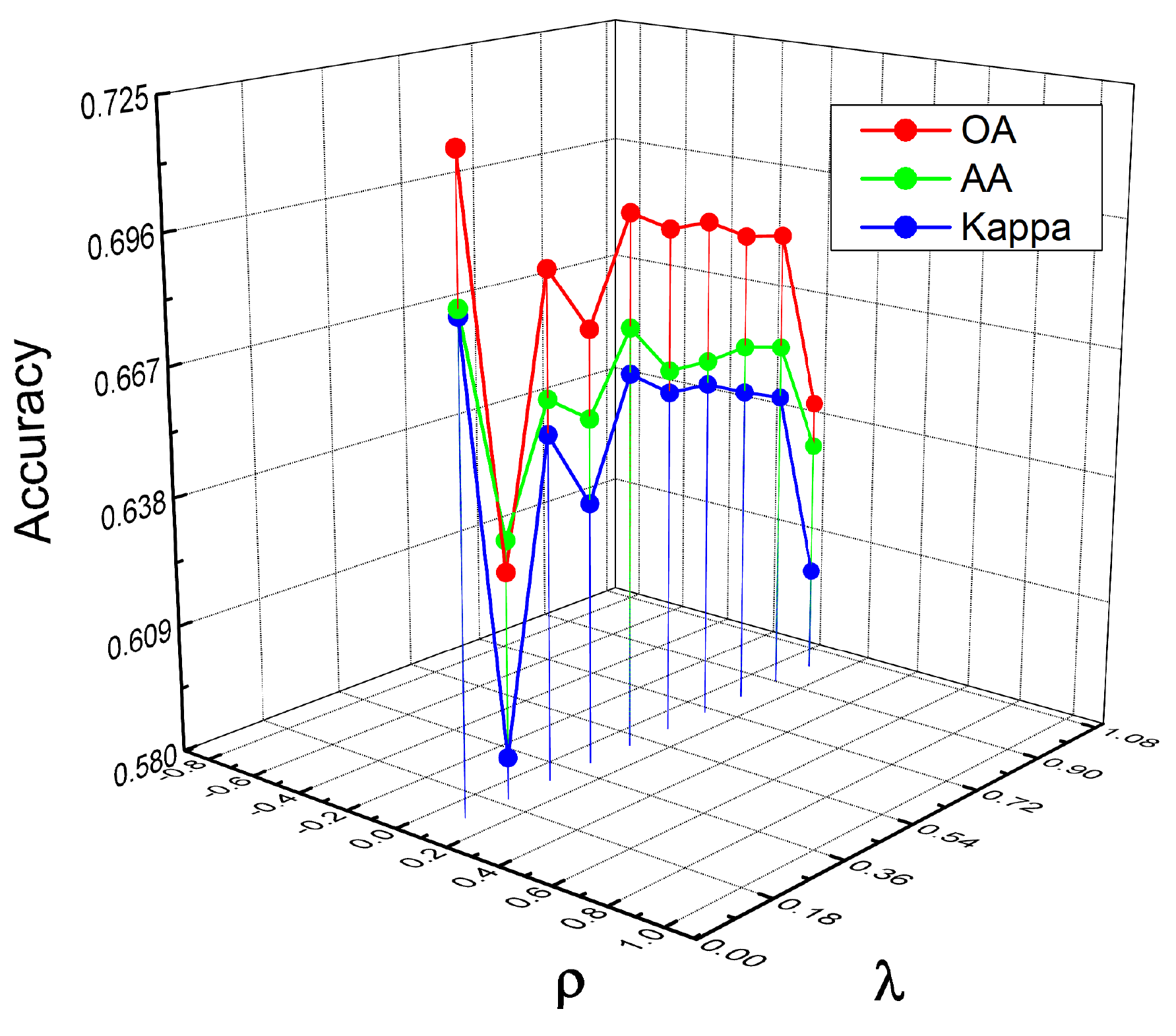}}
	\caption{Classification accuracy for Salinas using ADMM algorithm.(a)ADMM on different penalty parameters $\rho$; (b)ADMM on different regularization parameters $\lambda$}
	\label{Salinas admm}
\end{figure*}

Fixed $\rho$ or $\lambda$, these classification plots are shown in Fig.\ref{Pavia admm} and Fig.\ref{Salinas admm}, where the $x$-axis denotes the range of values for $\rho$, the $y$-axis denotes the range of values for $\lambda$ and the $z$-axis is the accuracy on the test set.
These classification curves fluctuates, which is difficult to find the global solution and determine the optimal parameters.
For ADMM algorithm, we still need to set up related parameters in advance.
The classification results of the proposed ASDN and ADMM are listed in Table \ref{Pavia result31} and Table \ref{Salinas result31}.
Our method still outperform than ADMM algorithm.

\subsection{Comparisons of different classifiers}

\begin{figure*}[htbp]
	\centering
	\subfigure[]{\includegraphics[width=0.8in,height=1.6in]{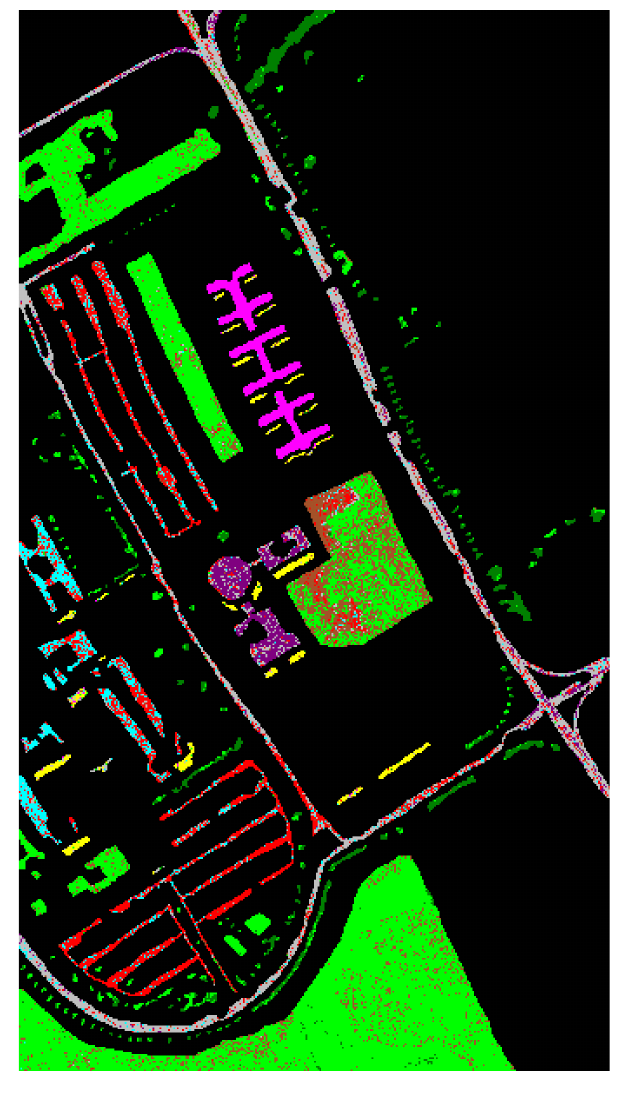}}
	\subfigure[]{\includegraphics[width=0.8in,height=1.6in]{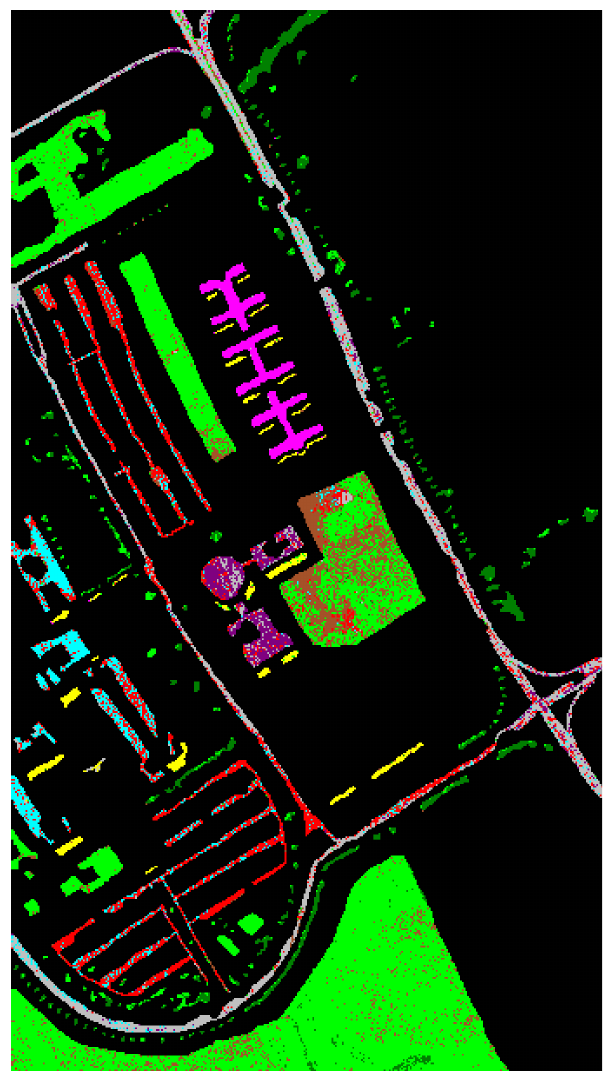}}
	\subfigure[]{\includegraphics[width=0.8in,height=1.6in]{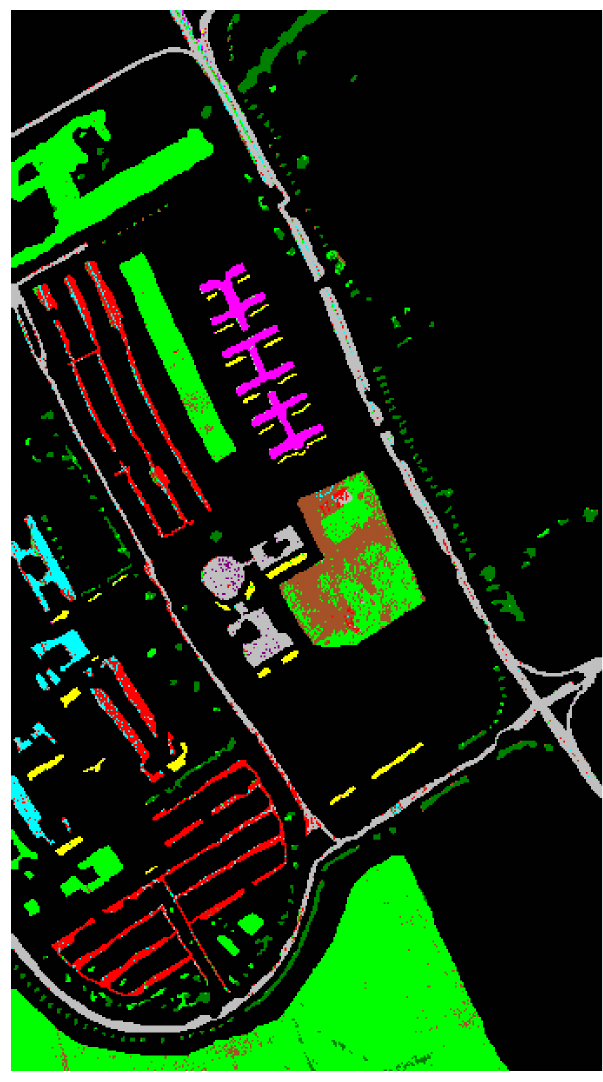}}
	\subfigure[]{\includegraphics[width=0.8in,height=1.6in]{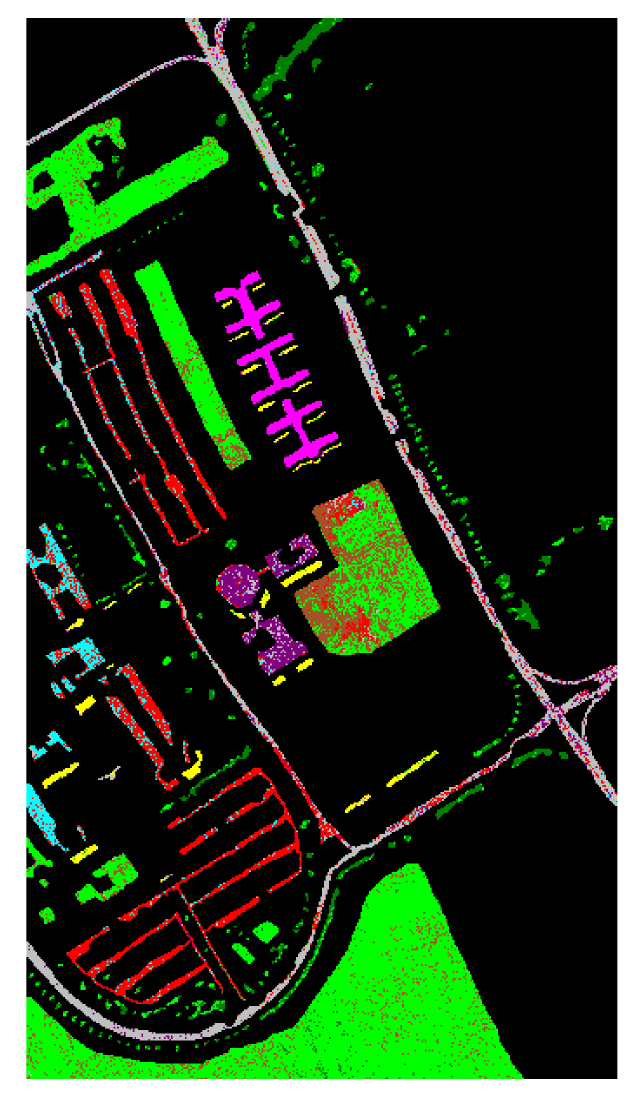}}
	\subfigure[]{\includegraphics[width=0.8in,height=1.6in]{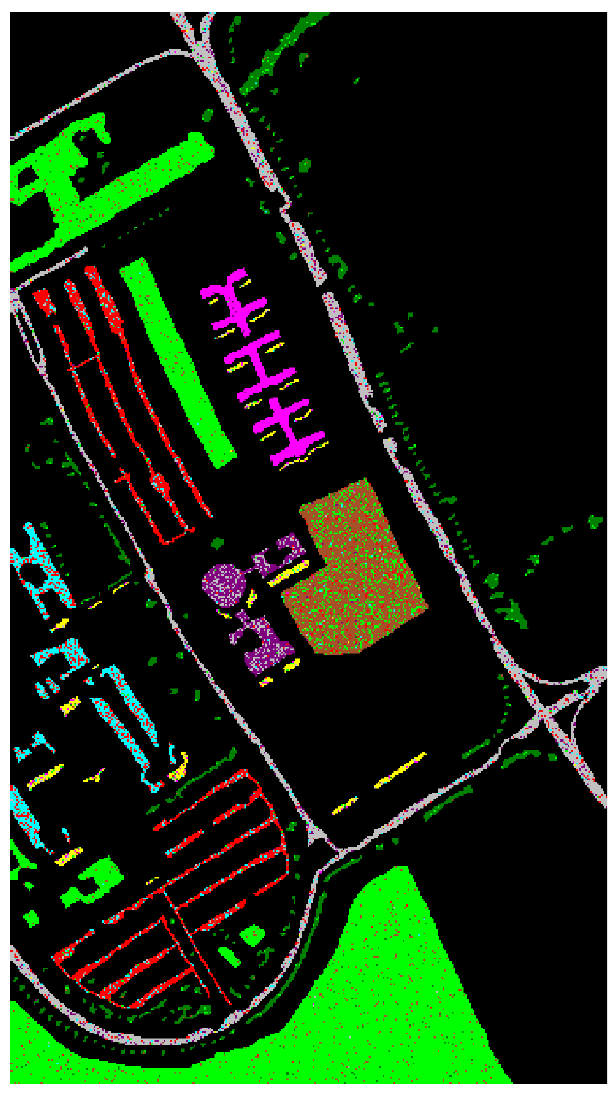}}
	\caption{Classification map for the Pavia University.(a)ELM; (b)KNN; (c)SVM;  (d)SAMP; (e)Adaptive Sparse Deep Network}
	\label{Pavia41}
\end{figure*}

\begin{figure*}[htbp]
	\centering
	\subfigure[]{\includegraphics[width=0.8in,height=1.6in]{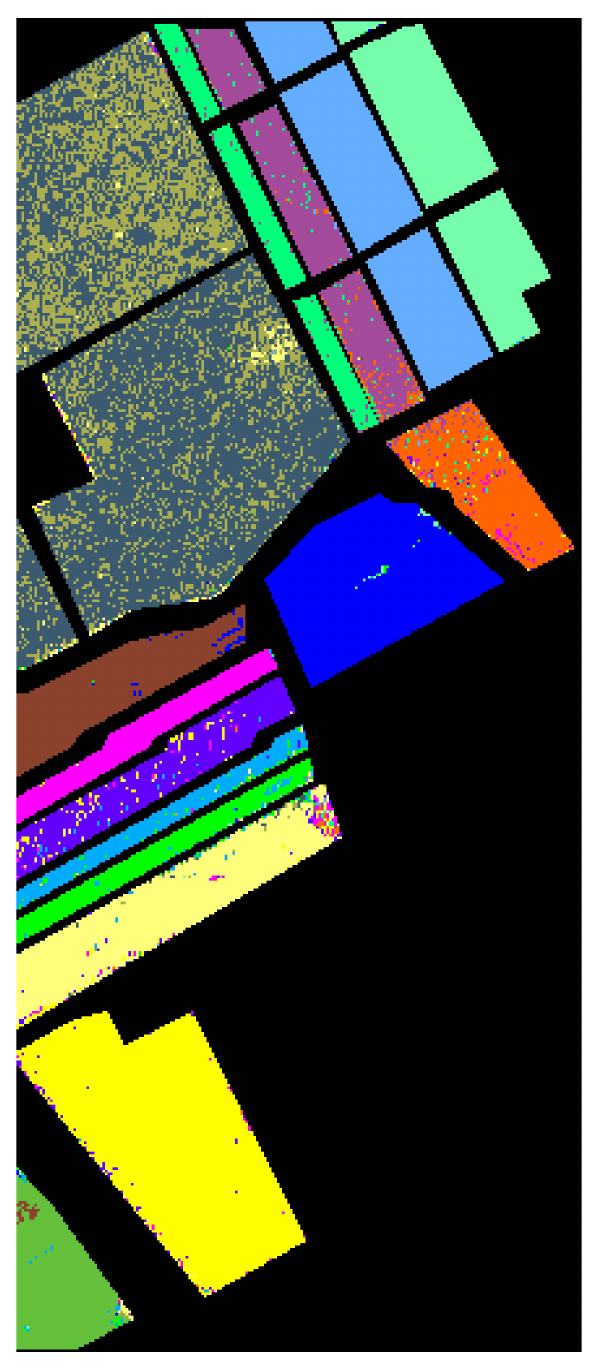}}
	\subfigure[]{\includegraphics[width=0.8in,height=1.6in]{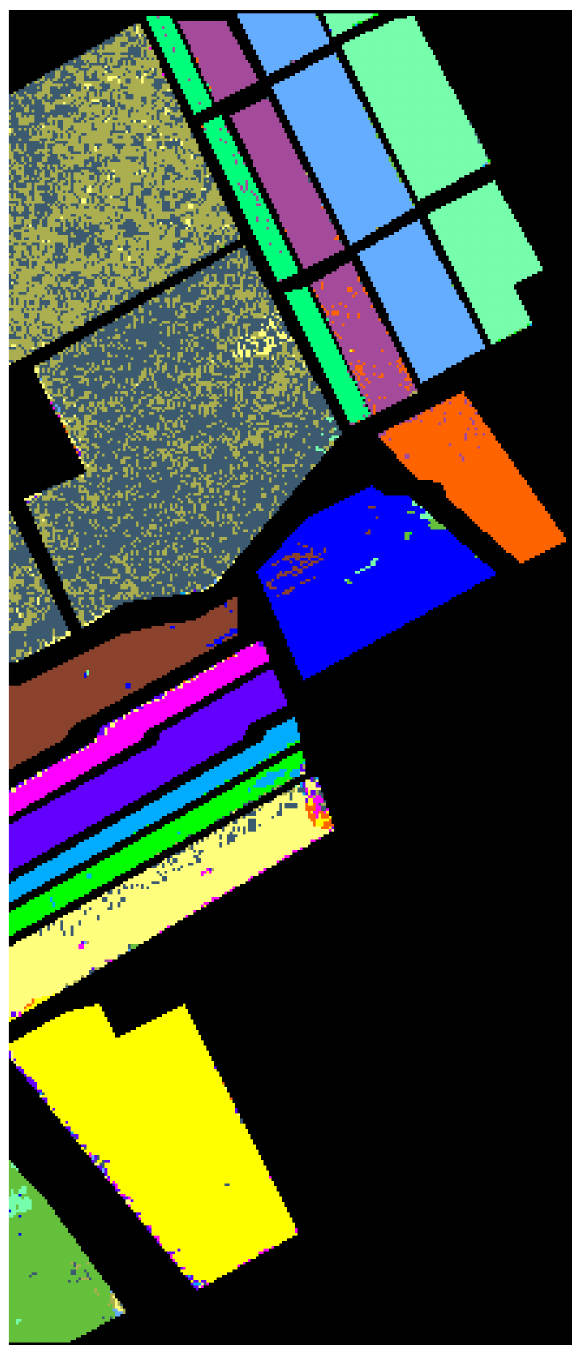}}
	\subfigure[]{\includegraphics[width=0.8in,height=1.6in]{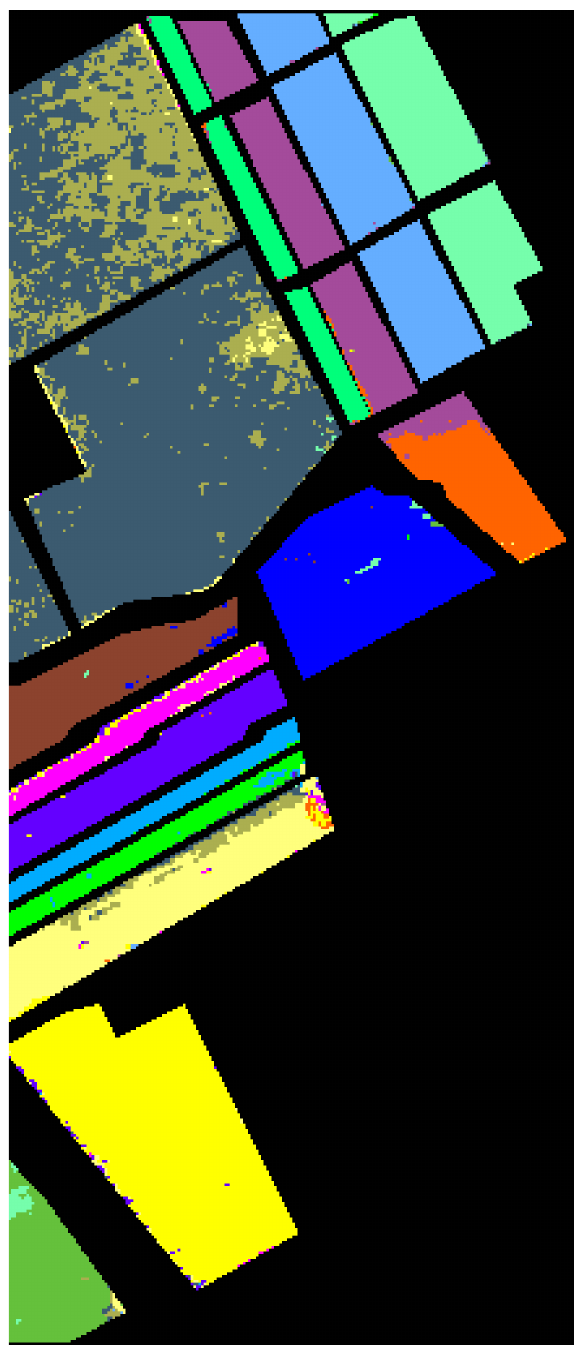}}
	\subfigure[]{\includegraphics[width=0.8in,height=1.6in]{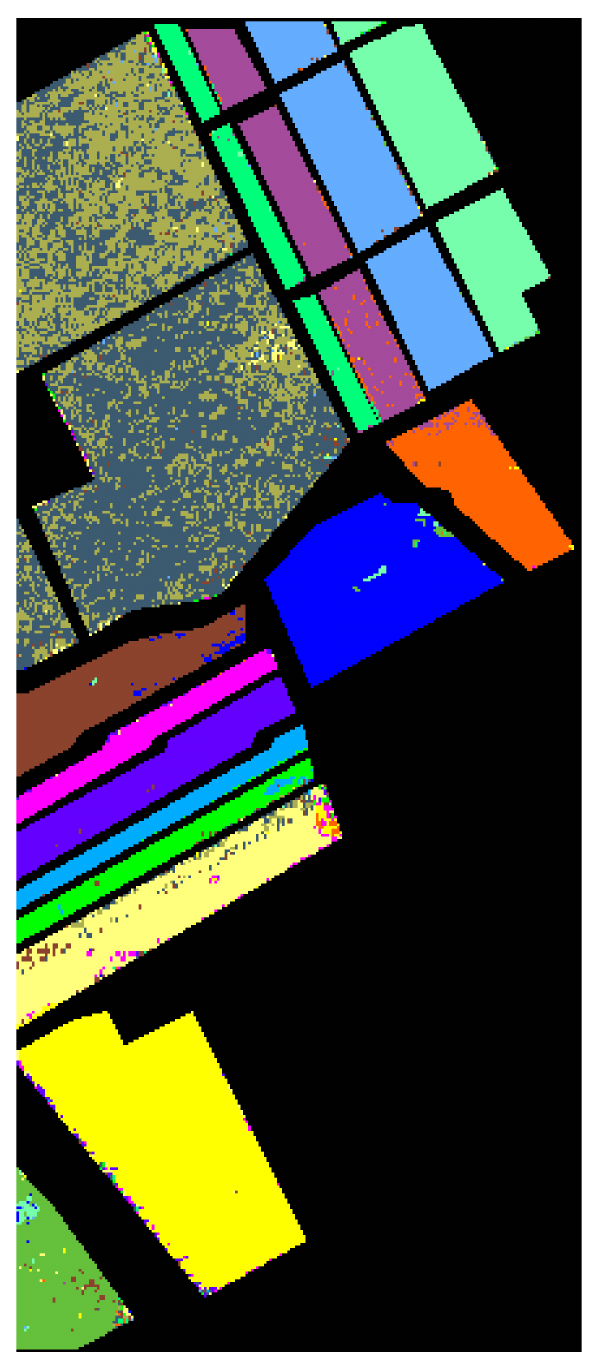}}
	\subfigure[]{\includegraphics[width=0.8in,height=1.6in]{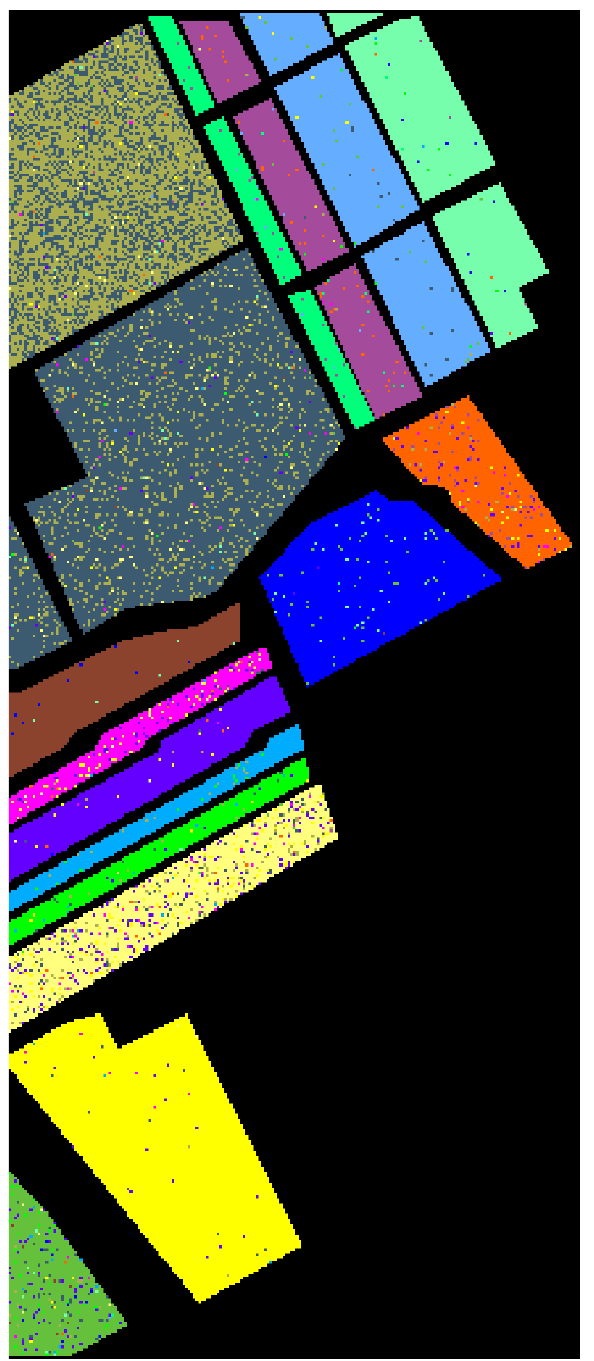}}
	\caption{Classification map for the Salinas. (a)ELM; (b)KNN; (c)SVM; (d)SAMP; (e)Adaptive Sparse Deep Network}
	\label{Salinas41}
\end{figure*}

In the four experiment, some traditional classifiers, like ELM~\cite{Miche2010OP}, KNN~\cite{Zhang2007ML}, SVM~\cite{Shevade2000Improvements}, sparsity adaptive matching pursuit (SAMP)~\cite{Do2009Sparsity}, were compared with our method ASDN.
The classification maps of the traditional classifiers are shown in Fig.\ref{Pavia41} and Fig.\ref{Salinas41}, and the results are summarized in Table \ref{Pavia result41} and Table \ref{Salinas result41}.
For KNN, it is difficult to train the model and interpret them.
The most innovative feature of the SAMP is its capability of signal reconstruction without prior information of the sparsity.
The SVM, the one-against-one strategy is applied.
In these classifiers, SVM yields the excellent overall performance.
But this classifier is still lower than the proposed ASDN, 1.4$\%$ in Pavia University and 1.54$\%$ in Salinas.
Therefore, our method is superior to these classifiers.

\begin{table}[!htbp]
	\renewcommand\arraystretch{1.1}
	\centering
	\caption{Classification accuracy ($\%$) for Pavia University using different classifiers}
	\begin{tabular}{|cccccc|}
		\hline
		\multirow{2}{*}{Class}  &\multirow{2}{*}{ELM}  &\multirow{2}{*}{KNN}  &\multirow{2}{*}{SVM}  &\multirow{2}{*}{SAMP}  &Adaptive Sparse \\
		&    &     &      &               &Deep Network       \\
		\hline 
		1	&69.38  &70.84  &87.25  &71.66   &76.35\\ 
		
		2	&91.50  &91.40  &96.30  &90.62   &95.33\\ 
		
		3	&49.07  &53.60  &49.48  &55.92   &65.11\\ 
		
		4	&81.56  &74.20  &86.24  &75.29   &\textbf{92.38}\\ 
		
		5	&95.70  &99.37  &98.06  &\textbf{98.57}  &98.93\\ 
		
		6	&58.38  &43.92  &53.72  &46.73   &\textbf{71.65}\\ 
		
		7	&40.03  &75.02  &38.36  &\textbf{71.58}   &67.25\\ 
		
		8	&59.66  &73.70  &80.22  &70.07     &77.33\\ 
		
		9	&89.01  &91.12  &90.59  &90.93    &68.9\\ 
		\hline 
		$OA$	&75.18  &77.76  &83.62  &77.62   &\textbf{85.02}\\ 
		\hline 
		$AA$	&70.47  &74.79  &75.58  &74.60    &\textbf{79.25}\\ 
		\hline 
		$k$	&66.57  &70.47  &77.79  &69.94   &\textbf{80.07}\\ 
		\hline 
	\end{tabular} 
	\label{Pavia result41}
\end{table}%

\begin{table}[htbp]
	\renewcommand\arraystretch{1.1}
	\centering
	\caption{Classification accuracy ($\%$) for Salinas using different classifiers}
	\begin{tabular}{|cccccc|}
		\hline
		\multirow{2}{*}{Class}   &\multirow{2}{*}{ELM} &\multirow{2}{*}{KNN} &\multirow{2}{*}{SVM} &\multirow{2}{*}{SAMP} &Adaptive Sparse \\
		&        &      &       &    &Deep Network       \\
		\hline 
		1	  &99.20 &98.14 &97.80 &98.59 &\textbf{99.23}\\ 
		
		2	  &\textbf{99.65}  &98.24 &98.60 &97.38 &97.58\\ 
		
		3	  &85.24 &92.71 &83.83 &\textbf{92.97} &91.51\\ 
		
		4	  &93.84  &\textbf{98.97} &98.23 &97.28 &97.69\\ 
		
		5	  &97.62  &94.90 &\textbf{97.63} &94.64 &97.26\\ 
		
		6	  &\textbf{99.65}  &99.58 &99.59 &99.27 &98.54\\ 
		
		7	  &\textbf{99.53}  &99.15 &99.35 &98.86 &99.16\\ 
		
		8	  &79.63  &68.47 &\textbf{88.16} &69.55 &85.76\\ 
		
		9	  &\textbf{99.36}  &97.88 &98.44 &96.91 &99.34\\ 
		
		10	&\textbf{92.25}  &88.09  &79.68  &85.76   &79.80\\ 
		
		11	&94.00  &\textbf{94.30}  &58.05  &94.09   &85.85\\
		
		12	&96.43  &99.77  &99.31  &\textbf{99.95}   &98.62\\
		
		13	&97.62  &\textbf{98.10}  &97.66  &97.41   &96.72\\
		
		14	&90.42  &87.44  &91.43  &91.64   &\textbf{96.05}\\
		
		15	&\textbf{60.42}  &60.23  &50.89  &59.76   &60.07\\
		
		16	&\textbf{95.32}  &89.60  &89.64  &88.26   &88.31\\      
		\hline 
		$OA$  &87.35  &85.56 &87.12 &85.37 &\textbf{88.66}\\ 
		\hline 
		$AA$  &91.52  &91.60 &89.28 &91.39 &\textbf{91.97}\\ 
		\hline 
		$k$	  &87.01  &83.94 &85.59 &83.73 &\textbf{87.34}\\ 
		\hline 
	\end{tabular} 
	\label{Salinas result41}
\end{table}%

\section{Conclusions}

In order to solve the problem that different of sparsity and regularization parameters has great influence on classification results in HSIs, the Adaptive Sparse Deep Network (ASDN) is presented in this paper, which is a deep architecture based on the data flow graph.
Data transfer and feedback are the basis of ASDN.
Based on ADMM algorithm, the deep data flow graph is built through the order of updating parameters in the sparse model.
The proposed ASDN consists three parts: input layer, hidden layer and output layer.
In Forward network, the under-sampled data from HSIs flows over the input layer to output layer in a order of updating parameters, and generate the optimal sparse vector.
Then for minimizing loss, parameters will be updating against by gradient computation in Back-Propagation network.
Therefore, in our method, related parameters are not manual setting in advance, which is different form other algorithms (greedy algorithms and $\ell_{1}$ algorithms) for solving this sparse model.
Experiment results indicate that our method outperforms than several traditional classifiers or other algorithms for sparse model.
In addition, we will make an attempt to extract new features from the HSIs for further improvements in experimental performance.

\bibliographystyle{IEEEtran}
\bibliography{reference}

\begin{thebibliography}{-------}
\providecommand{\natexlab}[1]{#1}

\bibitem[Benediktsson \em{et~al.}(2005)Benediktsson, Palmason, and
  Sveinsson]{Benediktsson2005Classification}
Benediktsson, J.A.; Palmason, J.A.; Sveinsson, J.R.
\newblock Classification of hyperspectral data from urban areas based on
  extended morphological profiles.
\newblock {\em IEEE Transactions on Geoscience \& Remote Sensing} {\bf 2005},
  {\em 43},~480--491.

\bibitem[Iyer \em{et~al.}(2017)Iyer, Raveendran, Bhuvana, and
  Kavitha]{Iyer2017Hyperspectral}
Iyer, R.P.; Raveendran, A.; Bhuvana, S.K.T.; Kavitha, R.
\newblock Hyperspectral image analysis techniques on remote sensing.
\newblock  Third International Conference on Sensing,  2017.

\bibitem[Tu \em{et~al.}({2019})Tu, Zhang, Kang, Zhang, and
  Li]{ISI:000460321300028}
Tu, B.; Zhang, X.; Kang, X.; Zhang, G.; Li, S.
\newblock {Density Peak-Based Noisy Label Detection for Hyperspectral Image
  Classification}.
\newblock {\em {IEEE TRANSACTIONS ON GEOSCIENCE AND REMOTE SENSING}} {\bf
  {2019}}, {\em {57}},~{1573--1584}.

\bibitem[Lin \em{et~al.}(2018)Lin, Li, Liu, and Li]{Lin2018Recent}
Lin, H.; Li, J.; Liu, C.; Li, S.
\newblock Recent Advances on Spectral-Spatial Hyperspectral Image
  Classification: An Overview and New Guidelines.
\newblock {\em IEEE Transactions on Geoscience \& Remote Sensing} {\bf 2018},
  {\em PP},~1--19.

\bibitem[Wang \em{et~al.}({2019})Wang, Wang, and Chen]{ISI:000477884900001}
Wang, A.; Wang, Y.; Chen, Y.
\newblock {Hyperspectral image classification based on convolutional neural
  network and random forest}.
\newblock {\em {REMOTE SENSING LETTERS}} {\bf {2019}}, {\em
  {10}},~{1086--1094}.

\bibitem[Donoho(2006)]{Donoho2006Compressed}
Donoho, D.L.
\newblock Compressed sensing.
\newblock {\em IEEE Transactions on Information Theory} {\bf 2006}, {\em
  52},~1289--1306.

\bibitem[Wang and Celik(2017)]{Wang2017Sparse}
Wang, H.; Celik, T.
\newblock Sparse Representation-Based Hyperspectral Data Processing: Lossy
  Compression.
\newblock {\em IEEE Journal of Selected Topics in Applied Earth Observations \&
  Remote Sensing} {\bf 2017}, {\em PP},~1--10.

\bibitem[Julien \em{et~al.}(2007)Julien, Michael, and
  Guillermo]{Julien2007Sparse}
Julien, M.; Michael, E.; Guillermo, S.
\newblock Sparse representation for color image restoration.
\newblock {\em IEEE Transactions on Image Processing} {\bf 2007}, {\em
  17},~53--69.

\bibitem[Michael and Michal(2006)]{Michael2006Image}
Michael, E.; Michal, A.
\newblock Image denoising via sparse and redundant representations over learned
  dictionaries.
\newblock {\em IEEE Tip} {\bf 2006}, {\em 15},~3736--3745.

\bibitem[Dong \em{et~al.}({2013})Dong, Zhang, Shi, and Li]{ISI:000318016600030}
Dong, W.; Zhang, L.; Shi, G.; Li, X.
\newblock {Nonlocally Centralized Sparse Representation for Image Restoration}.
\newblock {\em {IEEE TRANSACTIONS ON IMAGE PROCESSING}} {\bf {2013}}, {\em
  {22}},~{1618--1628}.

\bibitem[He \em{et~al.}({2017})He, Yi, Cheung, You, and
  Tang]{ISI:000395476200008}
He, Z.; Yi, S.; Cheung, Y.M.; You, X.; Tang, Y.Y.
\newblock {Robust Object Tracking via Key Patch Sparse Representation}.
\newblock {\em {IEEE TRANSACTIONS ON CYBERNETICS}} {\bf {2017}}, {\em
  {47}},~{354--364}.

\bibitem[John \em{et~al.}(2009)John, Yang, Arvind, S~Shankar, and
  Yi]{John2009Robust}
John, W.; Yang, A.Y.; Arvind, G.; S~Shankar, S.; Yi, M.
\newblock Robust face recognition via sparse representation.
\newblock {\em IEEE Transactions on Pattern Analysis \& Machine Intelligence}
  {\bf 2009}, {\em 31},~210--227.

\bibitem[Gao \em{et~al.}({2017})Gao, Ma, and Yuille]{ISI:000399396400035}
Gao, Y.; Ma, J.; Yuille, A.L.
\newblock {Semi-Supervised Sparse Representation Based Classification for Face
  Recognition With Insufficient Labeled Samples}.
\newblock {\em {IEEE TRANSACTIONS ON IMAGE PROCESSING}} {\bf {2017}}, {\em
  {26}},~{2545--2560}.

\bibitem[Wang \em{et~al.}({2019})Wang, He, and Li]{ISI:000456936500022}
Wang, Q.; He, X.; Li, X.
\newblock {Locality and Structure Regularized Low Rank Representation for
  Hyperspectral Image Classification}.
\newblock {\em {IEEE TRANSACTIONS ON GEOSCIENCE AND REMOTE SENSING}} {\bf
  {2019}}, {\em {57}},~{911--923}.

\bibitem[Yi \em{et~al.}(2011)Yi, Nasrabadi, and Tran]{Yi2011Hyperspectral}
Yi, C.; Nasrabadi, N.M.; Tran, T.D.
\newblock Hyperspectral Image Classification Using Dictionary-Based Sparse
  Representation.
\newblock {\em IEEE Transactions on Geoscience \& Remote Sensing} {\bf 2011},
  {\em 49},~3973--3985.

\bibitem[Zhang \em{et~al.}(2014)Zhang, Li, Huang, and Zhang]{Zhang2017A}
Zhang, H.; Li, J.; Huang, Y.; Zhang, L.
\newblock A Nonlocal Weighted Joint Sparse Representation Classification Method
  for Hyperspectral Imagery.
\newblock {\em IEEE Journal of Selected Topics in Applied Earth Observations \&
  Remote Sensing} {\bf 2014}, {\em 7},~2056--2065.

\bibitem[Baron \em{et~al.}(2012)Baron, Duarte, Wakin, Sarvotham, and
  Baraniuk]{Baron2012Distributed}
Baron, D.; Duarte, M.F.; Wakin, M.B.; Sarvotham, S.; Baraniuk, R.G.
\newblock Distributed Compressive Sensing {\bf 2012}.

\bibitem[Fang \em{et~al.}(2017)Fang, Wang, Li, and
  Benediktsson]{Fang2017Hyperspectral}
Fang, L.; Wang, C.; Li, S.; Benediktsson, J.A.
\newblock Hyperspectral Image Classification via Multiple-Feature-Based
  Adaptive Sparse Representation.
\newblock {\em IEEE Transactions on Instrumentation \& Measurement} {\bf 2017},
  {\em 66},~1646--1657.

\bibitem[Yan \em{et~al.}(2019)Yan, Chen, Zhai, Liu, and
  Liu]{ChenhdaHyperspectral}
Yan, J.; Chen, H.; Zhai, Y.; Liu, Y.; Liu, L.
\newblock Region-division-based joint sparse representation classification for
  hyperspectral images.
\newblock {\em IET Image Processing} {\bf 2019}, {\em 13},~1694--1704.

\bibitem[Wei \em{et~al.}(2016)Wei, Li, Fang, Kang, and
  Benediktsson]{Wei2016Hyperspectral}
Wei, F.; Li, S.; Fang, L.; Kang, X.; Benediktsson, J.A.
\newblock Hyperspectral Image Classification Via Shape-Adaptive Joint Sparse
  Representation.
\newblock {\em IEEE Journal of Selected Topics in Applied Earth Observations \&
  Remote Sensing} {\bf 2016}, {\em 9},~1--1.

\bibitem[Zhang \em{et~al.}(2017)Zhang, Xu, Yang, Li, and Zhang]{Zhang2017}
Zhang, Z.; Xu, Y.; Yang, J.; Li, X.; Zhang, D.
\newblock A Survey of Sparse Representation: Algorithms and Applications.
\newblock {\em IEEE Access} {\bf 2017}, {\em 3},~490--530.

\bibitem[Wang \em{et~al.}({2019})Wang, Zou, Tang, Li, and
  Shang]{ISI:000480665700027}
Wang, Y.; Zou, C.; Tang, Y.Y.; Li, L.; Shang, Z.
\newblock {Cauchy greedy algorithm for robust sparse recovery and multiclass
  classification}.
\newblock {\em {SIGNAL PROCESSING}} {\bf {2019}}, {\em {164}},~{284--294}.

\bibitem[Sayed \em{et~al.}({2019})Sayed, Hassanien, and
  Azar]{ISI:000457458000013}
Sayed, G.I.; Hassanien, A.E.; Azar, A.T.
\newblock {Feature selection via a novel chaotic crow search algorithm}.
\newblock {\em {NEURAL COMPUTING \& APPLICATIONS}} {\bf {2019}}, {\em
  {31}},~{171--188}.

\bibitem[Davenport and Wakin(2010)]{Davenport2010Analysis}
Davenport, M.A.; Wakin, M.B.
\newblock Analysis of Orthogonal Matching Pursuit Using the Restricted Isometry
  Property.
\newblock {\em IEEE Transactions on Information Theory} {\bf 2010}, {\em
  56},~4395--4401.

\bibitem[Dai and Milenkovic(2009)]{Dai2009Subspace}
Dai, W.; Milenkovic, O.
\newblock Subspace Pursuit for Compressive Sensing Signal Reconstruction.
\newblock {\em IEEE Transactions on Information Theory} {\bf 2009}, {\em
  55},~2230--2249.

\bibitem[Needell and Vershynin(2010)]{Needell2010Signal}
Needell, D.; Vershynin, R.
\newblock Signal Recovery From Incomplete and Inaccurate Measurements Via
  Regularized Orthogonal Matching Pursuit.
\newblock {\em IEEE Journal of Selected Topics in Signal Processing} {\bf
  2010}, {\em 4},~310--316.

\bibitem[Wang \em{et~al.}(2011)Wang, Kwon, and Shim]{Jian2011Generalized}
Wang, J.; Kwon, S.; Shim, B.
\newblock Generalized Orthogonal Matching Pursuit.
\newblock {\em IEEE Transactions on Signal Processing} {\bf 2011}, {\em
  60},~6202--6216.

\bibitem[Beck and Teboulle(2009)]{Beck2009A}
Beck, A.; Teboulle, M.
\newblock A Fast Iterative Shrinkage-Thresholding Algorithm for Linear Inverse
  Problems.
\newblock {\em Siam J Imaging Sciences} {\bf 2009}, {\em 2},~183--202.

\bibitem[Ye and Xie(2011)]{Ye2011Split}
Ye, G.B.; Xie, X.
\newblock Split Bregman method for large scale fused Lasso.
\newblock {\em Computational Statistics \& Data Analysis} {\bf 2011}, {\em
  55},~1552--1569.

\bibitem[Wright \em{et~al.}(2009)Wright, Nowak, and
  Figueiredo]{Wright2009Sparse}
Wright, S.J.; Nowak, R.D.; Figueiredo, M.A.T.
\newblock Sparse reconstruction by separable approximation.
\newblock {\em IEEE Transactions on Signal Processing} {\bf 2009}, {\em
  57},~2479--2493.

\bibitem[Boyd \em{et~al.}(2010)Boyd, Parikh, Chu, Peleato, and
  Eckstein]{Boyd2010Distributed}
Boyd, S.; Parikh, N.; Chu, E.; Peleato, B.; Eckstein, J.
\newblock Distributed Optimization and Statistical Learning via the Alternating
  Direction Method of Multipliers.
\newblock {\em Foundations \& Trends in Machine Learning} {\bf 2010}, {\em
  3},~1--122.

\bibitem[198(1983)]{1983}
{\em [Studies in Mathematics and Its Applications] Augmented Lagrangian
  Methods: Applications to the Numerical Solution of Boundary-Value Problems
  Volume 15 || Chapter 1 Augmented Lagrangian Methods in Quadratic
  Programming};  1983.

\bibitem[Fukushima(1992)]{Fukushima1992Application}
Fukushima, M.
\newblock Application of the alternating direction of multipliers to separable
  convex programming problems.
\newblock {\em Computational Optimization \& Applications} {\bf 1992}, {\em
  1},~93--111.

\bibitem[Spyridon~Kontogiorgis(1998)]{Spyridon1998A}
Spyridon~Kontogiorgis, R.R.M.
\newblock A Variable-Penalty Alternating Directions Method for Convex
  Optimization.
\newblock {\em Mathematical Programming} {\bf 1998}, {\em 83},~29--53.

\bibitem[Yang \em{et~al.}(Nov.2018)Yang, Sun, Huibin, and Xu]{YangADMM}
Yang, Y.; Sun, J.; Huibin, L.I.; Xu, Z.
\newblock ADMM-CSNet: A Deep Learning Approach for Image Compressive Sensing.
\newblock {\em IEEE Transactions on Pattern Analysis and Machine Intelligence}
  {\bf Nov.2018}, {\em PP},~1--1.

\bibitem[Luo \em{et~al.}(2018)Luo, Zou, Yao, Li, and Bai]{Luo2018HSI}
Luo, Y.; Zou, J.; Yao, C.; Li, T.; Bai, G.
\newblock HSI-CNN: A Novel Convolution Neural Network for Hyperspectral Image
  {\bf 2018}.

\bibitem[Slavkovikj \em{et~al.}(2015)Slavkovikj, Verstockt, Neve, Hoecke, and
  Walle]{Slavkovikj2015Hyperspectral}
Slavkovikj, V.; Verstockt, S.; Neve, W.D.; Hoecke, S.V.; Walle, R.V.D.
\newblock Hyperspectral Image Classification with Convolutional Neural
  Networks.
\newblock  2015.

\bibitem[Yuan \em{et~al.}({2019})Yuan, Zhang, Li, Shen, and
  Zhang]{ISI:000456936500044}
Yuan, Q.; Zhang, Q.; Li, J.; Shen, H.; Zhang, L.
\newblock {Hyperspectral Image Denoising Employing a Spatial-Spectral Deep
  Residual Convolutional Neural Network}.
\newblock {\em {IEEE TRANSACTIONS ON GEOSCIENCE AND REMOTE SENSING}} {\bf
  {2019}}, {\em {57}},~{1205--1218}.

\bibitem[Miche \em{et~al.}(2010)Miche, Sorjamaa, Bas, Jutten, and
  Lendasse]{Miche2010OP}
Miche, Y.; Sorjamaa, A.; Bas, P.; Jutten, C.; Lendasse, A.
\newblock OP-ELM: optimally pruned extreme learning machine.
\newblock {\em IEEE Transactions on Neural Networks} {\bf 2010}, {\em
  21},~158--162.

\bibitem[Zhang and Zhou(2007)]{Zhang2007ML}
Zhang, M.L.; Zhou, Z.H.
\newblock ML-KNN: A lazy learning approach to multi-label learning.
\newblock {\em Pattern Recognition} {\bf 2007}, {\em 40},~2038--2048.

\bibitem[Shevade \em{et~al.}(2000)Shevade, Keerthi, Bhattacharyya, and
  Murthy]{Shevade2000Improvements}
Shevade, S.K.; Keerthi, S.S.; Bhattacharyya, C.; Murthy, K.R.K.
\newblock Improvements to the SMO algorithm for SVM regression.
\newblock {\em IEEE Transactions on Neural Networks} {\bf 2000}, {\em
  11},~1188--1193.

\bibitem[Do \em{et~al.}(2009)Do, Gan, Nguyen, and Tran]{Do2009Sparsity}
Do, T.T.; Gan, L.; Nguyen, N.; Tran, T.D.
\newblock Sparsity adaptive matching pursuit algorithm for practical compressed
  sensing.
\newblock  Conference on Signals, Systems \& Computers,  2009.

\end{thebibliography}

\end{document}